\documentclass[envcountsect,envcountsame,draft]{llncs}

\usepackage[]{amsmath}
\usepackage[]{amssymb}
\usepackage[]{amsxtra}
\usepackage[utf8]{inputenc}
\usepackage[T1]{fontenc}
\usepackage{stix}
\usepackage{ifthen}
\usepackage[shortlabels,inline]{enumitem}
\usepackage{fancyhdr}
\usepackage{array}
\usepackage[normalem]{ulem}

\usepackage{graphicx}
\usepackage{tikz}
\usetikzlibrary{shapes.geometric}
\usetikzlibrary{chains}
\usepackage{wrapfig}

\usepackage{aliascnt}
\usepackage[
    final,
    unicode=true,
    pdftitle={First- and Second-Order Models of Recursive Arithmetics},
    pdfauthor={Paul J. Voda, Ján Kľuka},
    pdfkeywords={},
]{hyperref}

\AtBeginDocument{%
    \setcounter{tocdepth}{2}%
}
\makeatletter
\renewcommand{\subsubsection}{\@startsection
  {subsubsection}
  {2}
  {0mm}
  {-\baselineskip}
  {-0.5\baselineskip}
  {\normalfont\normalsize\bfseries}}
\makeatother

\numberwithin{subsubsection}{section}
\numberwithin{equation}{subsubsection}

\newcommand{\PAR}[2][\relax]{\subsubsection{\kern-0.5em{\boldmath#2.}}\if#1\relax\else\label{par:#1}\fi}

\newcommand{\ppsp}{\nobreak\hspace{.22222em plus.166667em minus.055555em}}

\newcommand{\XEL}[2]{(\ref{eq:#1\if #2\@\else:#2\fi})}

\newcommand{\XEG}[2]{\ref{par:#1}(\ref{eq:#1\if #2\@\else:#2\fi})}

\newcommand*\KERNABBRDOT{.\kern-0.08333em}

\newaliascnt{DEFINITION}{subsubsection}

\spnewtheorem{DEFINITION}[DEFINITION]{\definitionname}%
             {\normalfont\bfseries}{\normalfont}
\aliascntresetthe{DEFINITION}

\newaliascnt{THEOREM}{subsubsection}

\spnewtheorem{THEOREM}[THEOREM]{\theoremname}%
             {\normalfont\bfseries}{\normalfont\itshape}
\aliascntresetthe{THEOREM}

\newaliascnt{PROPOSITION}{subsubsection}

\spnewtheorem{PROPOSITION}[PROPOSITION]{\propositionname}%
             {\normalfont\bfseries}{\normalfont\itshape}
\aliascntresetthe{PROPOSITION}

\newaliascnt{LEMMA}{subsubsection}
\spnewtheorem{LEMMA}[LEMMA]{\lemmaname}%
             {\normalfont\bfseries}{\normalfont\itshape}
\aliascntresetthe{LEMMA}

\newaliascnt{COR}{subsubsection}

\spnewtheorem{COR}[COR]{\corollaryname}%
             {\normalfont\bfseries}{\normalfont\itshape}
\aliascntresetthe{COR}

\spnewtheorem*{CLAIM}{\claimname}%
             {\normalfont}{\normalfont\itshape}
\makeatletter
\def\@spbegintheorem#1#2#3#4{\trivlist
                 \item[\hskip\labelsep{#3#2\@thmcounterend}]{#3#1.\ }#4}
\def\@spopargbegintheorem#1#2#3#4#5{\trivlist
      \item[\hskip\labelsep{#4#2\@thmcounterend}]{#4#1. (#3)\ }#5}
\def\@thmcounterend{}
\makeatother

\DeclareMathAlphabet{\bmit}{OT1}{cmr}{bx}{it}

\newcommand{\UNFIN}{{\bf\sffamily\small UNFINISHED}:\ }

\newcommand{\RW}{\ensuremath{\rightarrow}}
\newcommand{\LW}{\ensuremath{\leftarrow}}
\newcommand{\LRW}{\ensuremath{\leftrightarrow}}
\newcommand{\ARR}{\rightarrowaccent}

\newcommand{\LA}{\ensuremath{\langle}}
\newcommand{\RA}{\ensuremath{\rangle}}

\newcommand{\PS}[1]{{\mid}#1{\mid}}

\newcommand{\pAND}[1]{\mathord#1\allowbreak\ }

\newcommand{\N}{\ensuremath{\mathbb{N}}}

\newcommand{\DOTMINUS}{\ensuremath{\mathbin{\text{\rlap{\kern.25em\raise.5ex\hbox{$\cdot$}}}{-}}}}

\newlength\lmindent\lmindent=0.5em
\makeatletter

\newenvironment{fleqn}{%
    \@fleqntrue%
    \@mathmargin = -1sp%
    \let\mathindent=\@mathmargin%
    \leftmargini = \lmindent%
    \@mathmargin\leftmargini minus\leftmargini%
}{}
\newenvironment{acl*}{\fleqn\start@align\z@\st@rredtrue\m@ne}{\endalign\endfleqn}

\newcommand\IND{\mathrm{I}}
\newcommand\COLL{\mathrm{B}}
\newcommand\COMPR{\mathrm{C}}

\newcommand\THEORY[1]{\mathsf{#1}}
\newcommand\RCA{\THEORY{RCA}_0}
\newcommand\WKL{\THEORY{WKL}}
\newcommand\PA{\THEORY{PA}}

\newcommand\INDAX{\mathrm{IND}}

\newcommand\LANG{\mathrm{L}}

\newcommand\STRUCT{\bgroup\STRUCTprime}
\newcommand\STRUCTprime[1][M]{\mathcal{#1}\egroup}
\newcommand\SETS[1][]{\mathcal{S}_{#1}}
\newcommand\DELTADEF{\Delta^0_1\text{-def}}

\newcommand{\COMP}{{\circ}}
\newcommand{\BASE}{\COMPR\Delta_0^0}

\newcommand{\BASIC}{\mathrm{BASIC}}

\newcommand{\TIME}{\mathbf{Time}}
\newcommand{\SPACE}{\mathbf{Space}}

\newcommand{\FN}{\mathcal{F}}

\newcommand\NEWALGEBRA[2]{%
    \newcommand{#1}[1][X]{#2\ifthenelse{\equal{##1}{\relax}}{}{({##1})}}%
}
\NEWALGEBRA{\DA}{\Delta}
\NEWALGEBRA{\GA}{{\cal A}}
\NEWALGEBRA{\TA}{{{\cal T}\mkern-4mu m}}
\NEWALGEBRA{\SA}{{{\cal S}\mkern-1.5mu p}}
\NEWALGEBRA{\DEA}{\Delta{\mathrm{e}}}
\NEWALGEBRA{\DSA}{\Delta\#}
\NEWALGEBRA{\SSA}{{{\cal S}\mkern-1.5mu p}\#}

\DeclareRobustCommand{\ARITHM}[2]{{#2[\relax]}\mathsf{RA}_{#1}}

\title{First- and Second-Order Models\\of Recursive Arithmetics}
\author{Ján Kľuka and Paul J. Voda}
\institute{First Draft May 8, 2017}

\begin{document}
\maketitle
\begin{abstract}
    We study a quadruple of interrelated subexponential subsystems of arithmetic
    WKL$_0^-$, RCA$^-_0$, I$\Delta_0$, and $\Delta$RA$_1$,
    which complement the similarly related quadruple
    WKL$_0$, RCA$_0$, I$\Sigma_1$, and PRA
    studied by Simpson,
    and the quadruple WKL$_0^\ast$, RCA$_0^\ast$,
    I$\Delta_0$(exp), and EFA studied by Simpson and Smith.

    We then explore the space of subexponential
    arithmetic theories between I$\Delta_0$ and I$\Delta_0$(exp).
    We introduce and study first- and second-order theories of recursive
    arithmetic $A$RA$_1$ and $A$RA$_2$
    capable of characterizing various computational complexity classes
    and based on function algebras~$A$, studied by Clote and others.
\end{abstract}

\section{Introduction}

Simpson in \cite{simpson:subsystems} studied the relations among four relatively weak subsystems of arithmetic:
$\WKL_0$, $\RCA$, $\IND\Sigma_1$, and
$\mathrm{PRA}$. The first two are second-order, the remaining  are first-order.
What is remarkable, is that all four theories share the same provably recursive functions, namely the primitive recursive ones.

Later Simpson and Smith \cite{simpson-smith:factorization} found the same kind of relationship among the weaker quadruple of theories:
$\WKL_0^\ast$, $\RCA^\ast$,
$\IND\Delta_0(\mathrm{exp})$, and $\mathrm{EFA}$ (Elementary Function Arithmetic) where all four share the Kalmar-elementary
functions as provably recursive.

The last four theories are discussed in a recent paper by Enayat and Wong \cite{enayat-wong}
as a framework  for the unification of model theory of the first and second-order arithmetic.
About the same time (summer of 2016) we have discovered even a weaker framework $\WKL_0^-$, $\RCA^-$,
$\IND\Delta_0$, and $\ARITHM1{\DA}$
(the theory of provably recursive functions of $\IND\Delta_0$).
The four theories are obtained by dropping the exponentiation
from the the four theories of Simpson and Smith. We have obtained the same kind of relationships among the theories.

This weakening is presented in sections \ref{sc:prelim} through \ref{sc:wkl} of this paper.
In order to obtain the result we had to be rather careful with the
coding of finite sequences, sets, and trees.
We have also decided to add the Cantor's pairing function to the language of arithmetic and to the basic axioms of PA
(Peano Arithmetic).

We do not treat the functions (in extensions of PA) in the usual way as denotations of $n$-ary function symbols, but
use the set-theoretical concept as sets of pairs (where the set-theoretical cartesian product is replaced by Cantor's pairing).
This is exactly
how the functions are treated in second-order subsystems of arithmetic. We have used this in our reformulation of
primitive recursive arithmetic $\mathrm{PRA}$ and the two above weaker systems of recursive arithmetics.
Instead of $n$-ary function symbols we have set constants (i.e. unary predicates). The advantage is that the respective
quadruples of theories share similar language and no rather awkward translation is needed (as in  \cite{simpson:subsystems}).

We can summarize our contributions as follows, The first one is the generalization
obtained by dropping the assumptions of exponentiation and proving similar results as above.
The second contribution is the introduction of first- and second- order theories of recursive arithmetics capable
of characterizations of various complexity classes. These theories are all subexponential
falling between the theories $\IND\Delta_0$ and $\IND\Delta_0(\mathrm{exp})$
and so our study of the interplay
of their first- and second-order model theories is an extension of the goal of Enayat and Wong.

Although it is possible to characterize the complexity classes in second-order arithmetic through set-comprehension axioms
(see e.g. \cite{zambella:poly,cook-kolokolova:poly}),
we have found
it simpler and more convenient to replace the set comprehension by function existence axioms.
This calls for the formulation of complexity classes as inductively defined function classes
(so called \emph{function algebras}, see \cite{clote}). A typical second-order function existence axiom is, for instance, the composition
axiom:
$\forall g,h{\in}\FN\exists f{\in}\FN\forall x\,f(x) = g(h(x))$.

We discuss the function algebras in \autoref{sc:algebras} where we discuss the ways of defining function algebras $\GA$.
The basic question we had to solve was how to specify the framework
for the definition of the operators of function algebras. We have tried several approaches and have finally settled for the use of
the Clausal Language (CL). CL is a subset of  PA (extended by definitions with functions) which we have developed
in 1997 and use in the teaching of computer programming and verification courses at our university.
Computer programming calls for a simple, readable, yet expressive language.
CL gives us a uniform treatment of complex recursive schemes
needed in operators of function algebras.

In \autoref{sc:first:order:ra} we assign to each function algebra $\GA$ a first-order theory $\ARITHM1{\GA}$,
called the recursive arithmetic of $\GA$
where we can talk about the functions of the algebra. The relationship between the two is like the one between the function algebra of
primitive recursive functions and the theory $\mathrm{PRA}$.

In \autoref{sc:provably:rec} we discuss the provably recursive functions of recursive arithmetics $\ARITHM1{\GA}$
which we show to be exactly the functions of the algebra $\GA$.

In \autoref{sec:ara2} we present the second order theories $\ARITHM2{\GA}$ extending the first-order theories $\ARITHM1{\GA}$ and
prove that they share the same provably recursive functions.

In \autoref{sc:complexity} we characterize some basic complexity classes by means of recursive arithmetics.

\section{Preliminaries}\label{sc:prelim}
When talking about the interplay  of first- and second-order arithmetic we have to decide on the often conflicting terminology and
notation. We generally prefer  Kaye's \cite{kaye} over Simpson's \cite{simpson:subsystems}. Since the pairing
function is central to the treatment of functions in this paper (all functions are unary, we do not have
any introduced functions symbols other than those mentioned
in the following paragraph), we have decided on one pairing function and added it to the basic symbols. This saves
the rather annoying constant referral to the theories extending $\LANG_1$ with pairing.

\PAR{Languages and Basic Axioms of our Fragments of First-Order Arithmetic}
The \emph{first-order language of arithmetic} $\LANG_1$ consists of the usual symbols $0$, $S$, $+$, $\cdot$, $<$ plus the binary
pairing function $(\cdot,\cdot)$ and its associated projection functions $H$, and $T$.

We use the
\emph{modified Cantor's pairing function} defined by:
\begin{gather}
 (x,y) = z \LRW 2 \cdot z = (x+y)\cdot(x+y+1) + 2 \cdot x + 2 \ .   \tag{$\mathrm{P0}$}
\end{gather}
This offsets the standard diagonal Cantor's function by one and makes it a bijection $\N^2 \mapsto \N \setminus\{0\}$.
The further properties of pairing are:
\begin{align}
 (x,y) = (x',y') & \RW x = x' \land y = y'  & z > 0 & \RW \exists x,y < z\, (x,y) = z  \ . \tag{$\mathrm{P1{-}2}$}
\end{align}
The symbols $H$ and $T$  are for the first (\emph{Head}) and second (\emph{Tail}) projection functions:
\begin{align}
 H((x,y)) & = x  & H(0)  & = 0  \tag{$\mathrm{P3{-}4}$} \\
 T((x,y)) & = y  & T(0)  & = 0 \ .  \tag{$\mathrm{P5{-}6}$}
\end{align}
We let pairing to associate to the right, i.e. $(a,b,c)$ abbreviates $(a,(b,c))$ and drop the unnecessary
parentheses in function applications involving pairing:
Thus $g\, h(x,y)$ abbreviates $g(h((x,y)))$.
This does not lead to confusion because apart from the six function symbols in $\LANG_1$ we never use other function symbols and
throughout the paper whenever we mention the term \emph{function} we mean a special set (see \autoref{par:functions}) and so
our functions are effectively unary.

The properties of the usual symbols of arithmetic are:
\begin{align}
  S(x) & \neq  0    &  S(x) =  S(y) & \RW x = y             \tag{$\mathrm{N}1{-}2$} \\
  x+0 & =  x        &  x +S(y) &=  S(x+y)                  \tag{$\mathrm{N}3{-}4$} \\
  x \cdot 0 & =  0  &  x \cdot S(y) & =  x \cdot y + x     \tag{$\mathrm{N}5{-}6$} \\
  x  & \not<  0     &  x < S(y) &  \LRW  x <y \lor x = y \tag{$\mathrm{N}7{-}8$} \\
  x < y \lor x = y & \lor y < x &   0 < x & \RW \exists y < x\,S(y) = x \ .\tag{$\mathrm{N}9{-}10$}
\end{align}
We designate the universal closures of the properties $\mathrm{N}1{-}10{+}\mathrm{P}0{-}6$ by $\BASIC$.

The reader will note that the group  $\mathrm{N}1{-}10$ deviates from the now standard basic axioms $\PA^-$ (see \cite{kaye})
and from the basic axioms of Simpson \cite{simpson:subsystems}.
We have decided on the axioms $\mathrm{N}1{-}9$ of Shoenfield \cite{shoenfield}.
The reasons behind the choice are that the axioms for $S$, $+$, $\cdot$ and $<$ are actually recurrences (as opposed to
the algebraic properties of $\PA^-$). The recurrences become important in our investigation in \autoref{sc:algebras} of
axiomatization of small fragments of arithmetic with
function symbols corresponding to the inductively defined classes of functions (think of primitive recursive functions and their
associated theory PRA). With our choice of $\LANG_1$ we, for instance, dispense with the annoying translation between the
standard language of arithmetic and that of PRA (see for instance  \cite{simpson:subsystems}).

The only extensions of the language of arithmetic $\LANG_1$ discussed in this paper are with set constants $\ARR{X}$
(unary predicate symbols). The languages are designated by $\LANG_1(\ARR{X})$.
We will almost always use the set constants
in the form $t \in X$ instead of the
predicate applications $X(t)$.

\PAR{Language of Second-Order Arithmetic}\label{par:lang:2}
The language for the theories of second-order arithmetic discussed in this paper is the two sorted
language $\LANG_2$ which extends $\LANG_1$
with variables ranging over sets and permits set quantification in formulas. For a set variable $X$ and a first-order term
$t$ the atomic formula $t \in X$ is in $\LANG_2$. We will use the set identity $X \subseteq Y$ as an abbreviation for
$\forall x(x \in X \RW x \in Y)$, $X=Y$ abbreviates $X \subseteq Y \land Y \subseteq X$,
and the relation $X < b$ is an abbreviation for $\forall x(x \in X \RW x < b)$. We often write
the last quantifier as $\forall x{\in}X\, x < b$.

\PAR{Structures for $\LANG_1$ and $\LANG_2$}
A structure for the language $\LANG_1$ of first-order arithmetic
is a tuple
$\STRUCT = (M, 0^{\STRUCT},\allowbreak {S}^{\STRUCT},\allowbreak
{+}^{\STRUCT},\allowbreak {\cdot}^{\STRUCT}, <^{\STRUCT},(\cdot,\cdot)^{\STRUCT},H^{\STRUCT},T^{\STRUCT})$.
We designate by $\N$ both the standard structure for $\LANG_1$ satisfying $\BASIC$ as well as its domain of natural numbers.
A structure for the language $\LANG_2$ of second-order arithmetic
$(\STRUCT, \SETS)$ with $\STRUCT$ a first-order structure for $\LANG_1$ and $\SETS \subseteq \mathcal{P}(M)$.
The set variables of $\LANG_2$ range over
the elements of $\SETS$.

The structures for the first order languages extended with set constants $\LANG_1(\ARR{X})$ are $(\STRUCT, \ARR{X})$ with
subsets of $\STRUCT$ assigned as meanings to the constants $\ARR{X}$ (note that we identify the constant symbols with their denotations).
Although such structures
look similar to the second-order structures our notation is a standard one
(see \cite{kossak:schmerl} page 3 for this treatment). We have the following obvious theorem:
\begin{PROPOSITION}\label{cl:first:second}
Let $(\STRUCT,\ARR{X})$ be a first-order structure for $\LANG_1(\ARR{X})$ and $(\STRUCT,\SETS)$ a second-order structure
(for $\LANG_2$) such that $\{\ARR{X}\} \subseteq \SETS$.
If $\varphi(\ARR{v})$ is formula of $\LANG_2$ without second-order quantifiers with all free (first and second-order) variables
among $\ARR{v}$ then for every $\ARR{v} \in \STRUCT \cup \{\ARR{X}\}$ we have
$(\STRUCT,\ARR{X}) \models \varphi(\ARR{v})$ iff $(\STRUCT,\SETS) \models \varphi(\ARR{v})$. \qed
\end{PROPOSITION}
The reader will note the subtlety that
the occurrence of the sequence $(\ARR{v})$ in the second-order satisfaction relation
is an assignment of values to variables of $\varphi(\ARR{v})$ whereas in the first-order satisfaction relation the set constants in $\ARR{v}$
replace the corresponding set variables in $\varphi$ (which thus becomes a formula in $\LANG_1(\ARR{X})$)
and only the first-order values from $\ARR{v}$ are in the assignment.

\PAR{Induction and Related Principles}\label{par:induction}
We designate the usual sets of arithmetical formulas $\Sigma_n$, $\Pi_n$, $\Delta_n$ without superscripts when they do not contain set
variables or constants, i.e. if they are in the languages $\LANG_1$. The same with superscripts, e.g, $\Sigma_n^0$, include
also the formulas with the set variables and constants. We will often call as arithmetical, also the formulas which are
only equivalent in some (usually implicitly understood)
structure or theory to a formula in the proper syntactic form. In addition to $\Delta_0$ formulas (with superscript or not), which are the
usual \emph{bounded} formulas,
we will designate a formula as $\Delta_1$ (possibly with superscript) only relatively to a structure
or a theory because
such a formula must satisfy the additional constraint:
\begin{gather*}
    \forall x(\varphi(x) \LRW \psi(x))       \tag{$\mathbf{\Delta}$}
\end{gather*}
with $\varphi(x) \in \smash{\Sigma_1}$ and $\psi(x) \in \smash{\Pi_1}$ (both possibly with superscripts).

For a formula $\varphi(x)$
we designate by  $\IND[\varphi(x)]$ the \emph{induction} formula:
\[
     \varphi(0) \land \forall x(\varphi(x) \RW \varphi(x+1))
    \RW \forall x\,\varphi(x) \ .
\]
We designate by $\COLL[\varphi(x,y)]$ the \emph{collection} formula:
\[
    \forall x{<}a\,\exists y\,\varphi(x,y) \RW
    \exists b\,\forall x{<}a\,\exists y{<}b\,\varphi(x,y)
\]
and by $\COMPR[\varphi(x)]$ the \emph{comprehension} formula:
\[
    \exists X\forall x ( x \in X \LRW \varphi(x) ) \ .
\]
In the comprehension formula $\varphi(x)$ may not contain $X$ as a parameter.
In all three kinds of formulas $\varphi$ may contain additional parameters.

When we call the three kinds of formulas
\emph{principles} (for example: the induction principles $\IND[\varphi(x)]$) then we understand the formulas to be universally closed.

For $\varGamma$ one of $\Sigma_1$, $\Delta_0$ (possibly with superscripts) we will designate by
$\IND[\varGamma]$, $\COLL[\varGamma]$, or $\COMPR[\varGamma]$ the sets of corresponding
induction, collection, or comprehension principles for $\varphi \in \varGamma$.

For $\varGamma$  one of $\Delta_0$, $\Delta_1$ (possibly with superscripts) we will designate by
$\IND[\varGamma]$ ($\COMPR[\varGamma]$) the sets of universal closures of
$\mathbf{\Delta} \RW \IND[\varphi]$ ($\mathbf{\Delta} \RW \COMPR[\varphi]$) for $\varphi \in \varGamma$.

\PAR{Fragments of Arithmetic with Limited Induction}
For $\varGamma$ one of $\Delta_0$, $\Delta_1$, $\Sigma_1$ (possibly with superscripts) we define the theory
$\IND\varGamma := \BASIC{+}\IND[\varGamma]$.

We define the theory  $\COLL\Sigma_1 := \IND\Delta_0{+} \COLL[\Sigma_1]$ (possibly with superscript).

For $\varGamma$ one of $\Delta^0_0$, $\Delta^0_1$ we define the second-order theory
$\COMPR\varGamma:=\BASIC{+} \INDAX{+} \COMPR[\varGamma]$ where
the \emph{induction axiom} $\INDAX$ is the universal closure of
\begin{align}
             0 \in X \land \forall x( x \in X \RW S(x) \in X ) \RW  x \in X \ .    \tag{$\INDAX$}
\end{align}

We say that a first-order theory $T$ in $\LANG_1(\ARR{X)}$ is \emph{inductive} if
the induction principles of $T$ hold also for the formulas containing the set constants $\ARR{X}$.

\section{The Second-Order Theory $\RCA^-$}
For the theory $\BASE$ we have:
\begin{PROPOSITION}\label{cl:base:ind}
\begin{enumerate}
\item
    $\BASE \vdash  \IND[\Delta^0_0]$,
\item
    $\BASE+ \COMPR[\Delta^0_1] \vdash  \IND[\Delta^0_1]$.
\end{enumerate}
\end{PROPOSITION}
\begin{proof}
1) Prove induction for $\varphi(x) \in \Delta^0_0$ by comprehension and then use $\INDAX$.

2) The same as 1) but with $\Delta^0_1$-comprehension.
\qed
\end{proof}

In the following we will strengthen the weak base theory $\BASE$ by axioms asserting existence of functions.

\PAR{Functions}\label{par:functions}
As mentioned above throughout this paper a ``function'' means a set acting like a function in the set-theoretical sense
where instead of set pairs $\LA x,y \RA$ we use pairing $(x,y)$. We use the symbols $f$, $g$, $h$, as set variables in
second-order contexts or as set (unary predicate) constants in first-order contexts.

Within theories extending $\BASE$ we define the property \emph{$f$ is a function}, in writing $f \in \FN$ as follows:
\begin{gather*}
  f \in \FN \LRW \forall x \exists y\,(x,y) \in f \land
        \forall x,y,y'((x,y) \in f \land (x,y') \in f \RW y = y') \land
          0 \not\in f\ .
\end{gather*}
We will often abbreviate $(x,y) \in f$ to $f(x) = y$. Function term $f(t)$ used in an atomic formula $\varphi(f(t))$ should be
understood as abbreviation for the \emph{unnested} form: $\exists y(f(t) = y \land \varphi(y))$
or $\forall y(y = f(t) \RW \varphi(y))$.

We use functions also in first-order theories in the languages $\LANG(\ARR{X})$ where $f \in \FN$ should
be viewed as a schema of abbreviations for the RHS of the above definition with the metavariable $f$ ranging over set constants.

The expression \emph{$f$ is a function in $\STRUCT[N]$} where $\STRUCT$ is a first- or second-order structure
means $f \in \FN^{\STRUCT[N]}$.

With $T$ being a first- or second-order theory we say that the function
 $f$ is \emph{polynomially bounded} if there is a term $t(x) \in \LANG_1$ such that $T$ proves
$\forall x,y(f(x) = y \RW y \leq t(x))$. The function $f$ is \emph{non-growing} if
$T \vdash \forall x,p,y(f(x,p) =y \RW y \leq p)$.

\PAR{Some Operators on Functions}\label{par:opers}
For the set variables $f$, $g$, and $h$,
we introduce
the following abbreviations (named on the right) as the universal closures
of the following formulas:
\begin{gather*}
   h(x) = z \land g(z) = y \RW f(x) = y  \tag{$f := g\COMP h$} \\
 {\setlength{\jot}{0pt}\begin{split}
  (g(y) = z  \RW {}&f(0,y) = z) \land {} \\
  (f(x,y) = v \land h((x,v),y) = w \RW {}&f(S(x),y) = w) \land  {} \\
   {}&f(0) = 0
 \end{split}} \tag{$f := \mathrm{PR}(g,h)$} \\
  f(0) = S(0) \land (f(x) = y \RW f S(x) = y + y)  \ .  \tag{$f:=\mathrm{E}$}
\end{gather*}
In the second-order context we will use a name of a function operator as a name of the axioms asserting the closure
under the operator:
\begin{gather}
\forall g,h {\in} \FN\exists f {\in} \FN\,f := g\COMP h    \tag{$\COMP$} \label{eq:composition2}\\
\forall g,h {\in} \FN\exists f {\in} \FN\,f  := \mathrm{PR}(g,h)   \tag{$\mathrm{PR}$} \label{eq:primrec2}\\
\exists f {\in} \FN\,f := \mathrm{E} \ .   \tag{$\mathrm{E}$}
    \label{eq:exp2}
\end{gather}
We will see below that the certain function operators are equivalent to set comprehension.
The first such equivalence is given in the following theorem:

\begin{THEOREM}\label{cl:comprehension}
Over $\BASE$ the theory $\COMPR[\Delta_1^0]$ is equivalent to the axiom $\COMP$.
\end{THEOREM}
\begin{proof}
$\BASE{+}\COMPR[\Delta_1^0] \vdash \COMP$: Take any $(\STRUCT,\SETS) \models \BASE{+}\COMPR[\Delta_1^0]$,
any functions $h,g \in \SETS$.
For all $w \in \STRUCT$ we have
\begin{align*}
 (\STRUCT,\SETS) \models & \exists z \exists x,y < w( w = (x,y) \land h(x) = z \land g(z) = y ) \LRW {} \\
   & \forall z\forall x,y < w( w = (x,y) \land h(x) = z \RW g(z) = y)
\end{align*}
and so by $\Delta_1^0$-comprehension with the LHS formula we obtain the desired set $f \in \SETS$ for which
\[ (\STRUCT,\SETS) \models  \FN(f) \land
    \forall x,y\bigl( (x,y) \in f  \LRW \exists z(h(x) = z \land g(z) = y \bigr) \ . \]

$\BASE{+}\COMP \vdash \COMPR[\Delta_1^0]$: Take any $(\STRUCT,\SETS) \models \BASE{+}\COMP$,
any $\Delta_0^0$-formulas $\varphi(\ARR{y},x)$, $\psi(\ARR{y},x)$,
any values of possible parameters $\ARR{v} \in \STRUCT\cup\SETS$ occurring in them, and
assume
$(\STRUCT,\SETS) \models \forall x\bigl(\exists \ARR{y}\,\varphi(\ARR{y},x) \LRW \forall \ARR{y}\,\psi(\ARR{y},x)\bigr)$.
We may assume w.l.o.g.\ that the variables $\ARR{y}$ are paired to a single one $y$.
We wish to find an $X \in \SETS$ such that $(\STRUCT,\SETS)$ satisfies $X = \{x \mid \exists y\,\varphi(x,y) \}$.

By $\Delta^0_0$-comprehension we obtain a set $h \in \SETS$ such that
\begin{align*}
  (\STRUCT,\SETS) \models \forall w\Bigl(
  w \in h \LRW \exists x,y < w\bigl(&w = (x,y,x) \land \psi(y,x) \RW \varphi(y,x) \land {} & \\
     & \forall z < y(\psi(z,x) \land \lnot \varphi(z,x))
  \bigr)
\Bigr)\ .
\end{align*}
We have
$(\STRUCT,\SETS) \models \forall x\exists y(\psi(y,x) \RW \varphi(y,x))$ because for any $x \in \STRUCT$ there is either a witness
to $\exists y\,\varphi(y,x)$ or a counterexample to $\forall y\,\psi(y,x)$. Thus $h$ is a function yielding pairs $(y,x)$.
By $\Delta^0_0$-comprehension we obtain a set $g \in \SETS$ such that
\begin{align*}
  (\STRUCT,\SETS) \models \forall w\Bigl(w \in g \LRW
  \exists x,y,z < w\bigl(& w = ((y,x),z) \land {} \\
                            & (\varphi(y,x) \RW z=1) \land  (\lnot \varphi(y,x) \RW z=0)\bigr)\Bigr) \ .
\end{align*}
Clearly, $g$ is the characteristic function of the formula $\varphi$. From $\COMP$ we
obtain $f = g\COMP h \in \SETS$  and for all $x \in \STRUCT$ we have
$(\STRUCT,\SETS) \models f(x) = 1 \LRW \exists y\,\varphi(x,y)$. Thus the desired $X$ is obtained by
$\Delta^0_0$-comprehension to satisfy $X = \{ x \mid f(x) = 1\}$.
\qed
\end{proof}

\PAR{The theory $\RCA^-$}\label{par:rca:minus}
We define $\RCA^- := \BASE{+}\COMP$.
In the view of \autoref{cl:comprehension}, the theory  $\RCA^-$ is equivalent to
$\COMPR\Delta^0_1$ and so by \autoref{cl:base:ind} $\RCA^- \vdash \IND[\Delta_1^0]$.

How does $\RCA^-$ compare to the theory $\RCA^\ast$ (see \cite{simpson-smith:factorization} or \cite{enayat-wong})
which is equivalent to $\IND\Delta_0^0(\mathrm{exp}){+}\COMPR[\Delta_1^0]$?
By the absence of the  axiom
\begin{align}
 \forall x \exists y\,2^x \doteq y \tag{$\mathrm{exp}$}
\end{align}
where $2^x \doteq y$ stands for a $\Delta_0$ formula defining the graph of exponentiation.
We, namely, have:
\begin{THEOREM}\label{cl:rca:ast}
Theories $\RCA^\ast$ and  $\RCA^-{+}\mathrm{E}$ are equivalent.
\end{THEOREM}
\begin{proof}
$\RCA^\ast \vdash \RCA^-{+}\mathrm{E}$: We work in $\RCA^\ast$ which proves $\INDAX$ and trivially
$\COMPR\Delta^0_0$. By \autoref{cl:comprehension} it proves
$\COMP$. We define $f := \{ v \mid \exists x{,}y{<}v( v = (x,y) \land 2^x \doteq y \}$ by $\Delta_0^0$ comprehension.
From $\mathrm{exp}$ we get $f \in \FN$, the recurrences in $f := \mathrm{E}$, and thus $\mathrm{E}$.

$\RCA^-{+}\mathrm{E} \vdash \RCA^\ast$: We work in $\RCA^-{+}\mathrm{E}$. From \autoref{cl:base:ind} we obtain $\IND[\Delta^0_0]$.
From \autoref{cl:comprehension} we get $\COMPR[\Delta^0_1]$. From $\mathrm{E}$ we get a function $f$. Since $f$ is a function, we
have
\[\forall x\bigl(\exists y( f(x) = y \land 2^x \doteq y) \LRW \forall y( f(x) = y \RW 2^x \doteq y)\bigr)  \]
and can use a $\Delta^0_1$ induction (which we obtain from \autoref{cl:base:ind})
to prove $\forall x\exists y( f(x) = y \land 2^x \doteq y)$ from which  $\mathrm{exp}$ directly follows.
\qed
\end{proof}

The theory $\RCA$ (see \cite{simpson:subsystems}) is defined as $\IND\Sigma_1{+}\COMPR[\Delta^0_1]$. Lemma 2.5 of
\cite{simpson-smith:factorization} asserts that $\RCA^\ast{+}\mathrm{PR}$ is equivalent to $\RCA$. The following
is a sharpening by dropping the exponentiation:
\begin{THEOREM}\label{cl:rca}
Theories $\RCA$ and
$\RCA^-{+}\mathrm{PR}$ are equivalent.
\end{THEOREM}
\begin{proof}
$\RCA \vdash \RCA^-{+}\mathrm{PR}$: $\RCA$ trivially proves $\COMPR[\Delta^0_0]$. It also proves the closure
under composition and primitive recursion (see \cite{simpson:subsystems}).

$\RCA^-{+}\mathrm{PR} \vdash \RCA$: We work in $\RCA^-{+}\mathrm{PR}$ and
define
\begin{align*}
  h := \bigl\{\, u \mid {} \exists x{,}y{,}v{<}u\bigl(& u = (((x,v),y),v+v) \lor
     u = ((0,y),0) \lor u = (0,0) \bigr) \,\bigr\}
\end{align*}
by $\COMPR[\Delta_0^0]$.
We obviously have $h \in \FN$. We define $g(y) = 1$ by $\COMPR[\Delta_0^0]$
and then  $f'(0,y) = g(y)$, $f'(S(x),y) = h((x,f'(x,y)),y)$ by primitive recursion. Finally, we define a function
$f(x) = f'(x,0)$ by $\COMPR[\Delta_0^0]$. $\Delta^0_0$-induction establishes $f := \mathrm{E}$ and hence $\mathrm{E}$.
By \autoref{cl:rca:ast} we get $\RCA^\ast$ and then use the above mentioned consequence of Lemma 2.5 of
\cite{simpson-smith:factorization}.
\qed
\end{proof}

\section{The First-Order Part of $\RCA^-$}
Simpson and Smith's \cite{simpson-smith:factorization} proved
that the first-order part of  $\RCA^\ast$ is $\IND\Delta_0(\mathrm{exp})+ \COLL[\Sigma_1]$.
In this section we will drop the exponentiation from their proof
and show that the first-order part of  $\RCA^-$ is $\COLL\Sigma_1$. This directly determines by \autoref{cl:cor:first:part:rca}
the first-order parts
not only of  $\RCA^\ast$ but also of $\RCA$ (see \cite{simpson:subsystems}).
Recall that the first-order part of a second-order theory $T_2$
is a first-order theory $T_1$ whose
theorems are identical to the theorems  of $T_2$ expressed in the language of $T_1$.

We have been inspired in \autoref{lemma:rca-=>bs1} by the unpublished proof of Gandy that over $\IND\Delta_0$
the least number principle for
$\Delta_1$ formulas implies $\COLL[\Sigma_1]$ (see \cite{slaman,hajek-pudlak:meta}).
This obviates the use of bounded recursion (needing exponentiation)
in the proof of Simpson and Smith. The structure of this
section is otherwise similar to the corresponding ones in
\cite{simpson:subsystems,simpson-smith:factorization}.

\begin{THEOREM}\label{lemma:rca-=>bs1}
    $\RCA^- \vdash \COLL[\Sigma^0_1]$.
\end{THEOREM}

\begin{proof}
We work in $\RCA^-$ and
take any $\smash{\Sigma^0_0}$ formula $\varphi(x,y,\ARR{z})$ possibly with number and set parameters $\ARR{w}$.
We wish to prove the principle $\COLL[\exists \ARR{z}\,\varphi(x,y,\ARR{z}]$. So we take any $a$, $\ARR{w}$, and assume
$\forall x {<} a \exists y \exists \ARR{z}\,\varphi(x,y,\ARR{z})$.
By taking $v:= \max(y,\ARR{z})$ we get
\begin{gather}
 \forall x {<} a\,\exists v\,\exists y{,}\ARR{z}{\leq} v\,\varphi(x,y,\ARR{z})
 \,. \tag{$\dagger$}
\end{gather}
Suppose we manage to obtain i) a function $f(x)$ yielding
the least bound $v$ and ii) we find its maximum $c:= \max_{z \in[0,\infty)}f(z)$.
Setting $b:=c+1$ we would then have
$\forall x{<}a\exists y{,}\ARR{z}{<} b$ $\varphi(x,y,\ARR{z})$ and we would get the desired conclusion of
the collection by dropping the bound on $\ARR{z}$.

Toward the goal i) we use $\Delta_0^0$ comprehension to define the set
\[ f = \{\, (x,v) \mid
    (x < a \land \exists y{,}\ARR{z}{\leq} v\,\varphi(x,y,\ARR{z}) \land
        \forall y{,} \ARR{z}{<} v\,\lnot\varphi(x,y,\ARR{z}) ) \lor
    (x \geq a \land v = 0)\,\} \ . \]
We wish to prove $f \in \FN$. That $v$ is unique is obvious. For the proof of its existence we take any $x$.
If $x \geq a$, we have $v:=0$. If $x < a$ we have $\exists y{,}\ARR{z}{\leq} v'\,\varphi(x,y,\ARR{z})$ for some $v'$ from $(\dagger)$
and by $\Delta_0^0$ least number principle we get the smallest such $v$ for which also $(x,v) \in f$ holds.

Toward the goal ii) suppose that we contrive to define the set
$X = \{\,x \mid f(x) = \max_{z \in [x,\infty)} f(z)\,\}$
Since $a \in X$, a $\Delta_0^0$ least number principle gives the least element~$m$ of $X$.
Furthermore, if we succeed in defining $Y = \{\,y \leq m \mid f(m) = \max_{z \in [y,m]} f(z)\}$
we will have $m \in Y$ and so $Y$ will have the least element $n$ for which we have $f(m) = \max_{z \in [n,\infty)}f(z)$.

Now, if $n > 0$, then $f(n \dotminus 1) > f(m)$ and we would get a contradiction $(n \dotminus 1) \in X$.
Thus $n=0$ and $c:=f(m)$ is the desired maximum of all $f(z)$.

It remains to define the sets $X$ and $Y$. The defining formula for $X$ is $\Sigma^0_1$ because it can be written as
\[ \exists v\bigl( f(x)=v \land (x < a \RW \forall z{\in}[x,a)\exists w\leq v\,f(z) = w )\bigr) \ .\]
This is equivalent to a  $\Pi^0_1$ formula:
$\forall v( f(x)=v \land x < a \RW \forall z{\in}[x,a)\exists w\leq v\,f(z) = w )$ and the
set $X$ is obtained
by $\Delta^0_1$ comprehension. The definition of $Y$ is similar. \qed
\end{proof}

\PAR{$\Delta^0_1$ Definability of Sets}\label{par:def:sets}
For a model $(\STRUCT,\SETS)$ for $\LANG_2$ we designate by
$\DELTADEF(\STRUCT,\SETS)$ the subset of ${\cal P}(\STRUCT)$ consisting of \emph{$\Delta^0_1$-definable} sets $X$, i.e.
such that there is a~$\Sigma^0_1$ formula~$\varphi(x,\ARR{v})$
and a $\Pi^0_1$ formula~$\psi(x,\ARR{v})$, possibly with \emph{parameters} $\ARR{v} \in \STRUCT \cup \SETS$,
such that $(\STRUCT,\SETS) \models \forall x\bigl(\varphi(x,\ARR{v}) \LRW \psi(x,\ARR{v}) \bigr)$,
and
\begin{gather}
 \text{for all $x \in \STRUCT$ we have $x \in X$ iff $(\STRUCT,\SETS) \models \varphi(x,\ARR{v})$.} \label{def:sets}
\end{gather}

\begin{LEMMA}\label{lemma:bs1=>rca-}
For every model $(\STRUCT,\SETS)$ for $\LANG_2$ we have
\[ \text{if $(\STRUCT,\SETS) \models \COLL\Sigma^0_1$ then $(M,\DELTADEF(\STRUCT,\SETS)) \models\RCA^-$} . \]
If the first model is countable, so is the expanded model.
\end{LEMMA}
\begin{proof}
Take a $(\STRUCT,\SETS) \models \COLL\Sigma^0_1$ and set $\SETS' := \smash{\DELTADEF}(\STRUCT,\SETS)$. We need to verify
that $(\STRUCT,\SETS')$
satisfies $\COMPR[\Delta^0_1]$ and $\INDAX$. For that we prove the auxiliary claim:
\begin{quote}\em
For every $\Delta^0_0$ formula $\theta$ possibly with parameters from $\STRUCT\cup \SETS'$ there is a $\Sigma^0_1$-formula
$\theta^\ast$
with parameters from $\STRUCT\cup \SETS$ such that $(\STRUCT,\SETS') \models \theta \LRW \theta^\ast$.
\end{quote}
The proof is by induction on the form of $\theta$ in the negation normal form.
If $\theta$ is $t \in X$ or $t \notin X$ with $X$ a set from $\SETS'$  $\Delta^0_1$ defined as in \autoref{par:def:sets}
we then define $\theta^\ast:=\varphi(t)$ or $\theta^\ast:=\lnot \psi(t)$ respectively.
In both cases $\theta^\ast$ is $\Sigma^0_1$ with parameters
from $\STRUCT\cup \SETS$.
The remaining
literals $\theta$ cannot have parameters from $\SETS'$ and we set $\theta^\ast := \theta$.
When $\theta$ is $\exists x < t\,\theta_1(x)$ then we set $\theta^\ast := \exists x < t\,\theta_1^\ast(x)$ which is
$\Sigma^0_1$ in  $(\STRUCT,\SETS')$.
When $\theta$ is a disjunction or conjunction we similarly obtain $\theta^\ast$ directly from IH.
The most interesting
case is when $\theta$ is $\forall x {<} t\,\theta_1(x)$. We can put $\theta_1^\ast(x)$ (which is without parameters in $\SETS'$)
into the form $\exists \ARR{z}\,\theta_2(x,\ARR{z})$ with $\theta_2 \in \Delta_0^0$. We then have
\begin{align*}
 (\STRUCT,\SETS') \models {} & \theta \LRW  \forall x {<} t\,\theta_1^\ast(x) \LRW
  \forall x {<} t\exists \ARR{z}\,\theta_2(x,\ARR{z}) \LRW {} \\
  & \forall x {<} t\exists y\exists \ARR{z}{\leq}y\,\theta_2(x,\ARR{z}) \LRW
   \exists b\forall x {<} t\exists y{<}b\exists \ARR{z} {\leq} y\,\theta_2(x,\ARR{z})
\end{align*}
where the last step is obtained in the direction $\RW$ from $\COLL[\Sigma^0_1]$
and in the direction $\LW$ by
predicate calculus. Thus we set $\theta^\ast$ to the last formula.
This ends the proof of the claim.

In order to prove $\COMPR[\Delta^0_1]$
assume $(\STRUCT,\SETS') \models \forall x\bigl(\exists y\,\varphi(x,y) \LRW \forall y\,\psi(x,y)\bigr)$
for $\smash{\Delta^0_0}$ formulas $\varphi$ and~$\psi$
possibly with parameters from~$M \cup \SETS'$ which we do not show. We assume that the possibly multiple quantifiers on $\ARR{y}$
have been contracted to $y$. We wish to show that
$X := \{ x \in \STRUCT \mid (\STRUCT,\SETS') \models \exists y\,\varphi(x,y)\}$ is in $\SETS'$. From the claim we get
\begin{align*}
(\STRUCT,\SETS') \models {} & \exists y\, \varphi^\ast(x,y) \LRW \exists y\,\varphi(x,y) \LRW \forall y\,\psi(x,y) \LRW {}\\
 & \lnot \exists y\,\lnot\psi(x,y) \LRW \lnot \exists y(\lnot\psi)^\ast(x,y) \LRW \forall y\lnot(\lnot\psi)^\ast(x,y)
\end{align*}
with the leftmost formula $\Sigma^0_1$ and the rightmost one $\Pi^0_1$ both with parameters at most from $\STRUCT\cup\SETS$.
Thus $(\STRUCT,\SETS) \models \exists y\, \varphi^\ast(x,y) \LRW \forall y\lnot(\lnot\psi)^\ast(x,y) $
and $X \in \DELTADEF(\STRUCT,\SETS)$.

In order to prove $(\STRUCT,\SETS') \models \INDAX$
we observe that $(\STRUCT,\SETS) \models \IND[\Delta^0_1]$ as the induction follows from $\COLL[\Sigma^0_1]$
(see \cite{slaman,hajek-pudlak:meta}).
We take any $X \in \SETS'$.
$X$ is $\Delta^0_1$ definable in $(\STRUCT,\SETS)$ and so there is a $\varphi(x)$
s.t. \autoref{def:sets} holds.
Hence $(\STRUCT,\SETS)$ satisfies the principle $\IND[\varphi]$ from which we get
that $(\STRUCT,\SETS')$ satisfies the principle $\IND[x \in X]$, i.e. $\INDAX$.

In order to finish the proof we observe that there are only countably many sets definable from countably many parameters.
\qed
\end{proof}

\begin{THEOREM}\label{theorem:rca-bs1}
Let $\STRUCT$ be a structure for $\LANG_1$. We have $\STRUCT \models \COLL\Sigma_1$ iff $(\STRUCT,\SETS) \models \RCA^-$
    for some $S \subseteq {\cal P}(\STRUCT)$.
\end{THEOREM}

\begin{proof}
The direction $\Rightarrow$ follows from \autoref{lemma:bs1=>rca-}.
In the direction $\Leftarrow$ if $(\STRUCT,\SETS)$ satisfies $\RCA^0$ then it also satisfies
$\IND[\Delta^0_0]$ by \autoref{cl:base:ind} and $\COLL[\Sigma^0_1]$ by \autoref{lemma:bs1=>rca-}. For the formulas $\LANG_1$ we thus
get that $\STRUCT$
satisfies $\BASIC{+}\IND[\Delta_0]{+}\COLL[\Sigma_1] =: \COLL\Sigma_1$.
\qed
\end{proof}

\begin{COR}\label{cl:cor:first:part:rca}
The first-order part of\/ $\RCA^-$ is\/ $\COLL\Sigma_1$,
\end{COR}

\begin{proof}
By Lemmas \ref{cl:base:ind} and \ref{lemma:rca-=>bs1} we have
    $\RCA^- \vdash \IND{\Delta_0},\COLL[\Sigma_1]$. Since $\COLL{\Sigma_1} = \IND\Delta_0{+}\COLL[\Sigma_1]$,
    $\RCA^-$ proves $\COLL\Sigma_1$.
    Vice versa, we need to show that  $\RCA^-$ is conservative over $\COLL\Sigma_1$ for sentences of $\LANG_1$.
    Thus we take any~$\varphi \in \LANG_1$
    not proved by $\COLL\Sigma_1$. This makes $\COLL\Sigma_1{+}\lnot \varphi$
    consistent and with a first-order model $\STRUCT$.
    By \autoref{theorem:rca-bs1},
    there is a model $(\STRUCT,\SETS)$ of~$\RCA^-$.
    Hence $\RCA^- \not \vdash \varphi$. \qed
\end{proof}

\section{Extension of $\RCA^-$ to $\WKL^-_0$}\label{sc:wkl}
In this section we follow the general development of \cite{simpson:subsystems,simpson-smith:factorization}
and expand models  of $\RCA^-$ to those of $\WKL^-_0$. Once again we have to be careful with the coding because of the
absence of exponentiation.
To mathematicians the details of coding are mostly immaterial. This is because they  work
in frameworks with primitive recursion (or at least with the exponentiation) available. The various
encodings of finite sets and sequences, trees, terms, and formulas are then invariant.
On the other hand, all feasible complexity classes are
subexponential. Hence the details of coding are usually relevant.

\PAR{Bounded Sets, $0{-}1$-Sequences, and Binary Trees}\label{par:codes:sets}
Within $\RCA^-$ we define
a set  $X$ \emph{bounded} iff $\exists b\,X < b$ (see \autoref{par:lang:2}).
The \emph{size} $\PS{X}$ of this set is the least such $b$. The graph of the size function $\PS{X} \doteq  b$
is $\Delta_0^0$ defined as a partial function.
We  use the \emph{Ackermann's} encoding of bounded sets with
the relation $x \in_{\text{Ack}} y$ holding iff the $x$-th least significant digit of the binary number representation of $y$ is $1$.
The relation has a $\Delta_0$ definition:
\[ x \in_{\text{Ack}} y \LRW \exists z,y_1 \leq y\, \exists y_2 <z
( 2^x \doteq z \land  y = (2\cdot  y_1 +1) \cdot z  + y_2) \ .
\]
If $X$ is bounded and $2^{\PS{X}}$ exists, i.e. if $\exists b {<} y(\PS{X} \doteq b \land 2^b \doteq y)$ for some $y$, then
the set $X$ is \emph{coded} by the number  $\sum_{i \in X}2^i < y$.

A finite sequence $\langle x_0,\ldots,x_{n-1} \rangle$ of length $n$, where
$\forall i {<} n\, x_i < 2$
is encoded by the number $(1x_0,\ldots,x_{n-1})_2$ which is the code of the set
$\{ i \mid i = n \lor (i < n \land x_{n-(i+1)} = 1)\}$ with the size $n+1$.
Thus every non-zero number codes a finite sequence where the empty sequence ($n=0$), is encoded by the number $1$,
the sequence $\langle 0100 \rangle$ by
the number $20 = (10100)_2$, the sequence $0^i$ of $i$ zeroes is encoded by $2^i$ if it exists, and the sequence $1^i$ by
$2^{i+1}-1$.
We will henceforth identify the finite sequences $\tau$ with their codes.

The (finite) sequence \emph{length} function $\PS{\tau}$ is defined as
\[ \PS{\tau} = y \LRW
           \exists p {\leq} \tau\exists x {<} p( 2^y \doteq p \land \tau = p+x) \lor \tau = 0 \land y = 0   \ . \]
The sequence \emph{concatenation} function $\sigma \star \tau$ is defined as
\[ \sigma \star \tau = \rho \LRW
           \exists p {\leq} \rho\exists x {<} p( 2^{\PS{\tau}} \doteq p \land \tau = p+x \land \sigma > 0 \land
                       \rho = \sigma\cdot p +x) \lor  \sigma \cdot \tau \cdot \rho  = 0 \land \rho = 0   \ . \]
The relation \emph{$\sigma$ is a subsequence of $\tau$}, in writing $\sigma \preceq \tau$, is $\Delta_0$ defined as
\[ \sigma \prec \tau \LRW
    \exists \rho {\leq} \tau(\sigma \star \rho = \tau \land \tau > 0)  \ . \]
The sequence \emph{$\sigma$ is a proper subsequence of $\tau$}, in writing $\sigma \prec \tau$ if, in addition to $\sigma \prec \tau$,
we have $\sigma < \tau$.

For a set $T$ we define the property of being \emph{$T$ is a (binary) tree}, in writing $T \in {\cal T} $, as follows:
\[ \text{$T$ is a tree} \LRW 0 \not\in T \land \forall \tau {\in} T\forall \sigma{\prec}\tau\,\sigma \in T \ . \]
Note that $1$ is the root of a tree $T \neq \emptyset$
and if $1 < \tau \in T$ then the parent of $\tau$ is $(\tau \div 2) \in T$ and $2\cdot \tau$ ($2\cdot \tau+1$)
is the left (right) child of $\tau$ neither necessarily in $T$ in which case $\tau$ is a leaf.

A tree $S$ is a {\em subtree} of the tree $T$ if $S \subseteq T$ .
A tree $P$ is a \emph{branch} if it is linearly ordered in $\prec$.
A tree $T$ is \emph{finite} if it is bounded
and \emph{infinite} otherwise. The property \emph{$T$ is an infinite tree} will be written as $T \in {\cal T}$.


\PAR{Monotone formulas}
Let $(\STRUCT,\SETS) \models \RCA^-$.
A  formula $\varphi(\tau,\ARR{v})$, possibly with parameters $\ARR{v}$, is \emph{monotone} in $\tau$ when for all
$\ARR{v} \in \STRUCT\cup\SETS$ we have
\begin{align*}
      (\STRUCT,\SETS) \models \forall \tau,\tau',\ARR{v}\bigl(\varphi(\tau,\ARR{v}) \land \tau \prec \tau'
          \RW \varphi(\tau',\ARR{v})\bigr)  \ .
\end{align*}

\PAR{The Theory $\WKL^-_0$}
Denote by $\WKL^-_0$ the theory $\RCA^-{+}\WKL$ where the sentence
\begin{gather}
 \forall T {\in} {\cal T}\exists P{\in}{\cal T}(\text{$P$ is a branch} \land P \subseteq T)
                                                      \tag{$\WKL$}
\end{gather}
is called the \emph{Weak K\"onig lemma}.

In the following we will  show that every countable $(\STRUCT,\SETS) \models \RCA^-$ can be expanded to a model
$(\STRUCT,\SETS') \models \WKL^-_0$.
This is done by refining the forcing-like argument from \cite{simpson-smith:factorization} where we add to $\SETS$
an infinite branch contained in every infinite tree in $\SETS$.

\PAR{Generic branches}
Let $\STRUCT[N]:= (\STRUCT,\SETS)$ be a model of $\RCA^-$.
A property ${\cal D} \subseteq {\cal T}^{\STRUCT[N]}$ of infinite trees is $\STRUCT[N]$-definable if there is a
formula  $\varphi(T,\ARR{v}) \in \LANG_2$ and parameters $\ARR{v} \in \STRUCT[N]$ such that
for every $T \in \SETS$ we have
\[ T \in {\cal D} \mathrel{\text{iff}} \STRUCT[N] \models T \in {\cal T} \land \varphi(T,\ARR{v}) \ . \]
Such a ${\cal D}$ is \emph{dense} if
$\STRUCT[N] \models \forall T {\in} {\cal T} \exists T'{\in}{\cal D}\,T' \subseteq T$.

A set $G \subseteq \STRUCT$ is a \emph{generic branch over infinite trees in $\SETS$} if for every dense definable
property ${\cal D}$
we have
\[ (\STRUCT,\SETS\cup G) \models \text{$G$ is an infinite branch} \land \exists T{\in}{\cal D}\  G \subseteq T \ . \]

\autoref{cl:generic:collection} and \autoref{cl:expansion}
are proved under the assumption that generic branches exist and
the \autoref{cl:generic} asserts that for countable structures they do.

\begin{LEMMA}\label{cl:generic:collection}
If $(\STRUCT,\SETS) \models \RCA^-$, $G$ is a generic branch over infinite trees in $\SETS$, and if
$\varphi(x,\tau)$ with parameters $\ARR{v}$ is monotone in $\tau$
then the expanded model $(\STRUCT,\SETS{\cup}\{G\})$ satisfies
the \emph{generic collection}:
\begin{align*}
  \forall a\bigl(\forall x < a\exists \tau \in G\,\varphi(x,\tau)
    \RW  \exists \tau \in G \forall x < a\, \varphi(x,\tau) \bigr)
\end{align*}
for all $\ARR{v} \in \STRUCT\cup\SETS$.
\end{LEMMA}
\begin{proof}
Abbreviate $\STRUCT[N] := (\STRUCT,S)$, $\STRUCT[N'] :=  (\STRUCT,\SETS{\cup}\{G\})$, and take a $\varphi(x,\tau)$ as in the
the theorem.
Take any $a,\ARR{v} \in \STRUCT[N]$, assume the hypothesis of the special collection, and
define the properties $\cal E$ and $\cal D$ of $T \in \SETS$:
\begin{align*}
  T \in {\cal E} & {} \mathrel{\text{iff}} \STRUCT[N] \models
                         {\cal T}(T) \land \exists x {<} a \forall \tau {\in} T \lnot \varphi(x,\tau) \\
  T \in {\cal D}  & {} \mathrel{\text{iff}} \STRUCT[N] \models
                   T \in {\cal E}  \lor
           \bigl( T \in {\cal T} \land \forall T' {\in} {\cal T}( T' \subseteq T \RW T' \notin {\cal E} \bigr) \ .
\end{align*}
The property $\cal D$ is dense because for any
$T \in {\cal T}^{\STRUCT[N]}$ when $\STRUCT[N] \models \exists T' {\in} {\cal T}(T'{\subseteq} T \RW T' \in {\cal E})$ then we
choose such a $T'$ to have the property $\cal D$. Otherwise there is no need to do anything
because we already have $T \in {\cal D}$.

Since $G$ is generic, there is a $T \in {\cal T}^{\STRUCT[N]}$ such that $T \in {\cal D}$ and $G \subseteq T$.
From the assumption we have $T \notin {\cal E}$.
We take any $x <^{\STRUCT} a$ and consider the set
\[ T' := \{ \tau \in \STRUCT \mid \STRUCT[N] \models \tau \in T \land \lnot \varphi(x,\tau)\} \ . \]
$T' \subseteq T$ is a tree by monotonicity of $\varphi$
and it is $\STRUCT[N]$-finite because otherwise we would have $T \in {\cal E}$.
Thus there is a $c_x \in \STRUCT$ such that
$\STRUCT[N] \models \forall \tau {\in} T(\tau> c_x \RW \varphi(x,\tau))$.
We have thus established:
\begin{gather}
 \STRUCT[N] \models \forall x {<}a \exists c \forall \tau {\in} T(\tau>c \RW \varphi(x,\tau) ) \ .\tag{$\dagger$}
\end{gather}
Our goal is to find an upper
bound of all $c_x$ for $x <^{\STRUCT} a$. We could use  $\Sigma_1^0$-collection in $\STRUCT[N]$
but for that we would need a suitable upper bound on $\tau$.
For reasons we will see below we take $4\cdot c+4$ as the bound and specialize
$(\dagger)$ to
  \[ \STRUCT[N] \models \forall x {<} a \exists c
         \forall \tau {\in} [c{+}1,4{\cdot}c{+}4)( \tau \in T \RW \varphi(x,\tau)) \ .
    \]
By applying collection we get a $b \in \STRUCT$
such that
\begin{gather} \STRUCT[N] \models \forall x {<} a \exists c < b \forall \tau {\in} [c{+}1,4{\cdot}c{+}4)
      ( \tau \in T \RW  \varphi(x,\tau)) \ . \tag{$\ddagger$}
\end{gather}
In order to prove the conclusion of the theorem we choose from the infinite $G$ a $\tau \in T$ such that
$\STRUCT[N'] \models \tau > 2{\cdot}b \land \tau \in G$.
We now take any $x <^{\STRUCT} a$ and use it in $(\ddagger)$ to obtain
a $c \in \STRUCT$ s.t.
\[ \STRUCT[N] \models c < b \land \forall \sigma{\in}[c{+}1,4{\cdot}c{+}4)( \sigma \in T \RW \varphi(x,\sigma)) \ . \]
We have
$\STRUCT[N] \models \PS{\tau} \geq \PS{2b} = \PS{b}+1 \geq \PS{c+1}+1$.
All sequences $\sigma \in \STRUCT$ s.t. $\STRUCT[N]$ satisfies $\PS{\sigma} = \PS{c+1}+1$ are such that
$\STRUCT [N] \models \sigma \in [2^{\PS{c+1}+1},2^{\PS{c+1}+2}) \subseteq [c+1,4\cdot c+4)$
and so we can choose one such that
$\STRUCT [N] \models \exists \rho(\tau = \sigma \star \rho \land \PS{\sigma} = \PS{c+1}+1)$.
But then
\[ \STRUCT[N] \models \sigma \preceq \tau  \land \sigma \in T \land
    \sigma \in[c+1,4{\cdot}c+4) \land \varphi(x,\sigma)\ . \]
We have $\STRUCT[N] \models \varphi(x,\tau)$
by monotonicity and hence $\STRUCT[N'] \models \varphi(x,\tau)$.
\qed
\end{proof}

\begin{THEOREM}\label{cl:expansion}
  If $(\STRUCT,\SETS) \models \RCA^-$ and  $G$ is a generic branch over infinite trees in $\SETS$ then
   $(\STRUCT,\SETS{\cup}\{G\})  \models \COLL\Sigma_1^0$\ .
\end{THEOREM}
\begin{proof}
Abbreviate $\STRUCT[N] := (\STRUCT,S)$ and $\STRUCT[N'] :=  (\STRUCT,\SETS{\cup}\{G\})$. We prove first a \emph{normal form}
property:
\begin{quote}
  \emph{
      For every $\Delta_0^0$-formula $\varphi(X)$ possibly with parameters $\ARR{v}$ there is
       a $\Delta_0^0$-formula $\bar{\varphi}(\tau)$ with the same parameters such that
       $\STRUCT[N]$ establishes its monotonicity in $\tau$
       and we have for all $\ARR{v} \in \STRUCT[N]$:
      \begin{align*}
        \STRUCT[N]'  \models \varphi(G) \LRW \exists \tau {\in} G\,\bar{\varphi}(\tau) \ .
      \end{align*}
  }
  \end{quote}
The proof is by induction on the form of $\varphi(X)$ in negation normal form
where we omit the straightforward proofs of monotonicity.
If $\varphi(X)$ is $t \in X$ then set $\bar \varphi(\tau) :\equiv t \preceq \tau$.
If $\varphi(X)$ is $t \not\in X$ then we observe that
$\STRUCT[N'] \models t \in G \LRW \forall \tau{\in}G(2\cdot t \leq \tau \RW t \prec \tau)$ and set
$\bar \varphi(\tau) :\equiv 2 \cdot t \leq \tau \land t \not\prec \tau$.
For the remaining
literals $\varphi$ we set $\bar \varphi(\tau) :\equiv \varphi(X)$ (the variable $X$ cannot occur in it).
For the compound formulas $\varphi(X)$
we obtain  the subformulas of $\bar \varphi(\tau)$ directly from IH.
When $\varphi(X)$ is of the form $\varphi_1(X) \land \varphi_2(X)$
we have from IH and  monotonicity:
  \[ \STRUCT[N'] \models \varphi_1(G) \land \varphi_2(G) \LRW \exists \tau_1 {\in} G\,\bar\varphi_1(\tau_1) \land
     \exists \tau_2 {\in} G\, \bar\varphi_2(\tau_2) \LRW \exists \tau {\in} G\, \bar \varphi(\tau) \]
and we set $\bar \varphi(\tau) :\equiv \bar\varphi_1(\tau) \land \bar\varphi_2(\tau)$.
The case when $\varphi(X)$ is a disjunction is similar and so is the case when $\varphi$
is $\exists x {<} t\,\psi(x,X)$ because we set $\bar \varphi(\tau) :\equiv \exists x{<} t\,\bar \psi(x,\tau)$ and by IH we have
     \[ \STRUCT[N'] \models  \exists x {<} t\,\psi(x,G) \LRW \exists x {<} t\exists \tau{\in} G\,\bar \psi(x,\tau)
         \LRW \exists \tau{\in} G\, \bar \varphi(\tau) \ .
         \]
The most interesting
case is when $\varphi(X)$
is $\forall x{<} t\,\psi(x,X)$.
From IH we have  $\STRUCT[N'] \models \varphi(G) \LRW \forall x {<} t\exists \tau {\in} G\,\bar \psi(x,\tau)$.
We set $\bar \varphi(\tau) :\equiv \forall x{<} t\,\bar \psi(x,\tau)$.
The implication $\STRUCT[N'] \models \varphi(G) \RW \exists \tau {\in} G\,\bar \varphi(\tau)$ follows from
\autoref{cl:generic:collection} and the converse from predicate calculus.
This ends the proof of the normal form property.

In order the prove the conclusion of the theorem it suffices to establish that $\STRUCT[N']$ satisfies $\COLL[\Sigma_1^0]$
and  $\IND[\Delta_0^0]$.

Take any $\smash{\Sigma^0_0}$ formula $\varphi(x,y,\ARR{z},X)$ possibly with number and set parameters $\ARR{w}$.
We wish $\STRUCT[N']$ to satisfy $\COLL[\exists \ARR{z}\,\varphi(x,y,\ARR{z},G)]$.
So take any $a,\ARR{w} \in \STRUCT[N]$, and assume
$\STRUCT[N'] \models \forall x {<} a \exists y \exists \ARR{z}\,\varphi(x,y,\ARR{z},G)$.
From the normal form property we get
\[ \STRUCT[N'] \models
         \forall x {<} a \exists y{,}\ARR{z}\exists \tau' {\in} G\,\bar \varphi(x,y,\ARR{z},\tau')
          \ .\]
Since $G$ is infinite, there is for any $x <^{\STRUCT} a$ a sequence $\tau \in G$
s.t. $\STRUCT[N'] \models y,\ARR{z} \leq \tau \land \tau' \preceq \tau$.
Using monotonicity we get
\begin{gather*}
 \STRUCT[N'] \models \forall x {<} a \exists \tau {\in} G \exists y{,}\ARR{z}{\leq} \tau\,\bar \varphi(x,y,\ARR{z},\tau)\ .
\end{gather*}
We now apply \autoref{cl:generic:collection} to obtain:
\begin{gather*}
 \STRUCT[N'] \models \exists \tau{\in}G \forall x {<} a \exists y{,}\ARR{z}{\leq} \tau\,\bar \varphi(x,y,\ARR{z},\tau)\ .
\end{gather*}
From this we get after some easy manipulation with $b:=\tau+1$:
\begin{gather*}
 \STRUCT[N'] \models \exists b \forall x {<} a \exists y{<}b \exists \ARR{z}\exists \tau{\in}G\,\bar \varphi(x,y,\ARR{z},\tau)
\end{gather*}
and it remains to apply the normal form property backwards to get
\begin{gather*}
 \STRUCT[N'] \models \exists b \forall x {<} a \exists y{<}b \exists \ARR{z}\,\varphi(x,y,\ARR{z},G)
\end{gather*}
as desired.

For the proof of $\IND[\Delta_0^0]$ we take a $\Delta_0^0$ formula $\varphi(x,X)$ possibly with parameters $\ARR{v}$.
We take any $\ARR{v} \in \STRUCT[N]$ and assume by way of contradiction
\begin{gather}
 \STRUCT[N'] \models \varphi(0,G) \land \forall x(\varphi(x,G) \RW \varphi(x+1,G)) \land \lnot \varphi(a,G) \tag{$\dagger$}
\end{gather}
for some $a \in \STRUCT$.

For every $x \in \STRUCT$ we have from the normal form property:
\[ \STRUCT[N'] \models  \exists \tau {\in} G\, \bar \varphi (x,\tau)  \mathrel{\text{iff}}
   \STRUCT[N'] \models  \varphi(x,G) \mathrel{\text{iff}}
   \STRUCT[N'] \models \forall \tau {\in} G\, \lnot \overline{\lnot \varphi}(x,\tau) \ .
\]
Thus $\STRUCT[N'] \models \forall x \leq a\exists \tau {\in} G
         (\bar \varphi (x,\tau) \lor \overline{\lnot \varphi} (x,\tau))$
and by $\Sigma^0_1$ collection (with $a:=a+1$) we get for some $b \in \STRUCT$:
\[ \STRUCT[N'] \models \forall x \leq a \exists \tau {<} b
         \bigl(\tau \in G \land (\bar \varphi (x,\tau) \lor \overline{\lnot \varphi}(x,\tau))\bigr) \ . \]
$G$ is an infinite tree, and so there is a $\sigma \in G$ such that $\sigma >^{\STRUCT}  b$
and for any $x \leq^{\STRUCT} a$ we get a $\tau <^{\STRUCT} b$, $\tau \in G$ such that
$\STRUCT \models \bar \varphi (x,\tau) \lor \overline{\lnot \varphi} (x,\tau)$.
As $G$ is a branch we have $\tau \prec^{\STRUCT} \sigma$ and from the monotonicity we obtain
$\STRUCT[N] \models \bar \varphi (x,\sigma) \lor \overline{\lnot \varphi}(x,\sigma)$.
Thus
\[ \STRUCT[N'] \models \varphi(x,G) \Rightarrow \STRUCT[N] \models \lnot \overline{\lnot \varphi}(x,\sigma) \Rightarrow
   \STRUCT[N] \models \bar \varphi (x,\sigma) \Rightarrow \STRUCT[N'] \models \varphi(x,G) \ . \]
From $(\dagger)$ we have $\STRUCT[N] \models \overline{\lnot \varphi}(a,\sigma)$
and by the least number principle in $\STRUCT[N]$ there is a least such $m \leq^{\STRUCT} a$. It cannot
be the case that $m = 0^{\STRUCT}$ and so
$\STRUCT[N] \models \bar \varphi(m \dotminus 1,\sigma) \land \overline{\lnot \varphi}(m,\sigma)$
contradicting  $(\dagger)$.
\qed
\end{proof}

\begin{LEMMA}\label{cl:generic}
Let $(\STRUCT,\SETS)$ be a countable model of $\RCA^-$. For every infinite tree $T \in \SETS$ there is
a generic branch $G$  over infinite trees in $\SETS$ such that $G \subseteq T$.
\end{LEMMA}
\begin{proof}
Abbreviate $\STRUCT[N] := (\STRUCT,\SETS)$ and enumerate all $\STRUCT[N]$-definable (with parameters) dense
sets into a countable sequence $\{ {\cal D}_i\}_{i \in \N}$. 

For every $b \in \STRUCT$ define
\[ T \in {\cal E}_b \mathrel{\text{iff}}  \STRUCT[N] \models T \in {\cal T} \land
  \exists \tau{\in}T( \PS{\tau} = b \land \forall \sigma{\in}T (\PS{\sigma} = b \RW \sigma = \tau)) \ .
\]
The sets ${\cal E}_b$ are
dense because given an infinite tree $T$ there must be a sequence $\tau \in  \STRUCT[N]$ s.t.
\[ \STRUCT[N] \models  \tau \in T \land \PS{\tau} = b \land \exists T' {\in} {\cal T}( T' \subseteq T \land \tau \in T') \ . \]
We form by $\Delta_0^0$-comprehension an infinite tree $T' \subseteq T$ such that
\[ \STRUCT[N] \models \sigma \in T' \LRW \sigma \in T \land ( \PS{\sigma} < \PS{\tau} \lor
   \exists \rho {\leq} \sigma\,\sigma = \tau \star \rho ) \ .
\]
We clearly have $T' \in {\cal E}_b$.

Given an infinite tree $T \in \SETS$, we set $T_0 := T$ and for $i \in \N$ we set $T_{i+1}$ to a ${\cal D}_i$ dense infinite tree
obtained for $T_i$. Clearly, for all $i \in \N$ we have $T_i \in {\cal T}^{\STRUCT[N]}$, $T_{i+1} \subseteq T_i$, and
the set $G := \bigcap_{i \in \N} T_i \subseteq T \subseteq \STRUCT$ is
an infinite branch because at every level $b \in \STRUCT$
it has exactly one sequence
and all of them are $\prec$ comparable.
For this the order of ${\cal E}_b$ in the enumeration is irrelevant, although for different orders the infinite
branches $G$ may differ.
Moreover, $G$ is generic because every dense definable
set must be ${\cal D}_i$ for some $i \in \N$ and we have $G \subseteq T_{i+1} \in {\cal D}_i$.
\qed
\end{proof}

\begin{THEOREM}\label{thm:rca-=>wkl-}
  Every countable  structure $(\STRUCT,\SETS) \models \RCA^-$ can be expanded to a
  countable structure $(\STRUCT,\SETS') \models \WKL^-_0$ with $\SETS \subseteq \SETS'$.
\end{THEOREM}
\begin{proof}
We will define a sequence of sets $\{ \SETS[i] \}_{i \in \N}$ such that for all $i,j \in \N$, $i < j$
we will have $\SETS[i] \subseteq \SETS[j] \subseteq \mathrm{P}(\STRUCT)$.

For that we need a function $\mathrm{Branch}_{a,b} := G$ for $G$ obtained by \autoref{cl:generic} with
$(\STRUCT,\SETS[(a,b){\dotminus}1])$ and an $T_b \in \SETS[a]$.
Here $T_b$ is an infinite tree at the $b$-th position
in some fixed enumeration of infinite trees in ${\cal T}^{(\STRUCT,\SETS[a])}$.

The sets $\SETS[i]$ are defined by
$\SETS[0] := \SETS$, and $\SETS[i+1]:= \DELTADEF(\STRUCT, \SETS[i]\cup \{\mathrm{Branch}_{a,b}\})$ where $a,b$ are such that $i+1 = (a,b)$.

Complete induction on $i$ establishes
\begin{quote}
\emph{
 $(\STRUCT,\SETS[i]) \models \RCA^-$ and if $i = (a,b)$ then there is  an infinite branch
$\mathrm{Branch}_{a,b}$ with $\mathrm{Branch}_{a,b} \subseteq T_b$
where $T_b$ is the  $b$-th tree $T_b$ in ${\cal T}^{(\STRUCT,\SETS[a])}$. We have $\mathrm{Branch}_{a,b},T_b \in \SETS[i]$.
}
\end{quote}
Indeed, there is nothing to prove when $i=0$. Otherwise we have $i = (a,b)$ for some $a,b < i$ and the
structures $(\STRUCT,\SETS[a])$ and $(\STRUCT,\SETS[i \dotminus 1])$ both satisfy $\RCA^-$
by IH. We use the last structure and the infinite tree $T_b \in \SETS[a] \subseteq \SETS[i \dotminus 1]$  in \autoref{cl:generic}
to obtain a generic branch $\mathrm{Branch}_{a,b} \subseteq T_b$. The structure $(\STRUCT,S_{i \dotminus 1}\cup \{\mathrm{Branch}_{a,b}\})$
satisfies $\COLL\Sigma^0_1$ by \autoref{cl:expansion} and so closing it by $\Sigma^0_1$-definitions by \autoref{lemma:bs1=>rca-}
yields $(\STRUCT,\SETS[i]) \models \RCA^-$ with $\mathrm{Branch}_{a,b},T_b \in \SETS[i]$.

We now set $\SETS^\ast := \bigcup_{i\in \N} \SETS[i]$ and claim that the structure
$\STRUCT[N] := (M,\SETS^\ast)$ is the desired countable structure extending $\SETS$  and satisfying $\WKL_0$.
The extension is trivial: $\SETS = \SETS[0] \subseteq \SETS^\ast$.
That the structure is countable, follows from the fact that it is the result of countably many operations which change a countable
structure to another countable one.
In order to establish that $\STRUCT[N] \models \RCA^-$,
it suffices to show that $\STRUCT[N] \models \INDAX,\COMPR{\Delta^0_0},\COMP$. For $\INDAX$ we take a set $X \in \SETS^\ast$.
It appears first in some $S_i$ and $(\STRUCT,\SETS[i])$ satisfies $\INDAX$. Similarly two functions $g,h \in \SETS^\ast$ appear
both in some $\SETS[i]$ which is closed under composition. The principle $\COMPR[\varphi]$ for a $\Delta_0^0$ formula
$\varphi$ is also similar because
all its set parameters must appear in some $\SETS[i]$ because there is only finitely many of them.
Thus already $(\STRUCT,\SETS[i])$ contains the comprehended set.

For $\STRUCT[N] \models \WKL$
we take a $T \in {\cal T}^{\STRUCT[N]}$. It appears for the first time at the $b$-th position in some
${\cal T}^{(\STRUCT,\SETS[a])}$. Thus $\mathrm{Branch}_{a,b}$ is an infinite branch in $T$ which is in $\SETS[(a,b)] \subseteq \SETS^\ast$.
\qed
\end{proof}

\begin{THEOREM}\label{cl:wkl:pi:2}
  $\WKL^-_0$ is $\Pi^1_1$ conservative over $\RCA^-$.
\end{THEOREM}
\begin{proof}
  Suppose that $\forall X\,\varphi(X) \in \Pi^1_1$ is not provable in  $\RCA^-$.
  Thus there is
  a countable model $(M,S) \models \RCA^-+ \exists X\lnot \varphi(X)$. Take $X \in S$ s.t.
  $(M,S) \models \lnot \varphi(X)$
  Expand the model to $(M,S') \models \WKL^-_0$. Since $S \subseteq S'$, we have $(M,S') \models \lnot \varphi(X)$.
  Thus $\WKL^-_0 \not\vdash \forall X\,\varphi(X)$. \qed
\end{proof}

\begin{COR}\label{cl:cor:first:part:wkl}
   The first-order part of $\WKL_0$ is the same as that of $\RCA^-$, namely $\COLL\Sigma_1$.
\end{COR}
\begin{proof}
   $\WKL_0$ is an extension of $\RCA^-$ so it proves all of the latter's theorems in $\LANG_1$. Vice versa,
   take any sentence $\varphi \in \LANG_1$ such that $\WKL_0^- \vdash \varphi$. Thus $\RCA^-\vdash \varphi$ by \autoref{cl:wkl:pi:2}
   because $\varphi$ is trivially $\Pi^1_1$. Thus $\WKL_0$ and $\RCA^-$ have the same theorems in $\LANG_1$
   which by \autoref{cl:cor:first:part:rca}
   are exactly the theorems of $\COLL\Sigma_1$.
    \qed
\end{proof}

\section{Function Algebras}\label{sc:algebras}

In this section we introduce operators for defining classes of
functions over natural numbers by inductive definitions. The classes
are called \emph{function algebras} by Clote in \cite{clote} where the reader will find a comprehensive overview of defining
classes of functions of computational complexity. We assign to every function algebra $\GA$ a first-order theory
$\ARITHM1{\GA}$ called the \emph{recursive arithmetic} of $\GA$.
This is similar to the going from the class of primitive recursive functions to the theory PRA
(Primitive Recursive Arithmetic) in the form presented in Simpson \cite{simpson:subsystems}.

However, rather than treating the functions as $n$-ary, we work with their pair contractions into unary
functions.
We have opted for this approach because of its direct connection to the second-order theories of recursive
arithmetics which will be discussed in the next section.

\PAR{Function Algebras}\label{par:operators}
A \emph{function operator} $f := \mathrm{op}(g_1,\ldots,g_n)$
is a mapping that takes $n \geq 0$
functions $g_1$, \ldots, $g_n$ in~$\N$
and yields a unique function~$f$ in~$\N$.
The \emph{oracle operator} $f := X_\ast$
is a mapping that given any set $X \subseteq \N$
yields the unique function~$f$ such that $(\N,X,f)$
satisfies
\begin{gather}
  \forall x\bigl((x \in X \RW f(x) = 1) \land
                 (x \not\in X \RW f(x) = 0)\bigr) \ .
  \tag{$f := X_\ast$}
\end{gather}
Clearly, $(\N,X,f) \models f \in \FN$.

For every \emph{oracle} $X \subseteq \N$
and a $k$-tuple of function operators~$\GA[\relax]$
a \emph{function algebra $\GA$} is the least set
that contains the function $f := X_\ast$
and is closed under the operators of~$\GA[\relax]$.
We view $\GA$ without $X$ specified as the \emph{class of algebras}
$\{ \GA \mid X \subseteq \N \}$.

Henceforth, every function  operator $f := \mathrm{op}(g_1,\ldots,g_n)$
will be specified by a formula in $\LANG_2$ with no free first-order variables, no-second-order quantifiers, and which
contains exactly the set variables $f$, $g_1$, \ldots, $g_n$.
The second-order set variables are to be viewed in first-order contexts as meta-variables ranging over the set
constants. This effectively turns operators into schemas. We require
that any structure $(\N,X,g_1,\ldots,g_n) \models g_1,\ldots,g_n \in \FN$
can be uniquely expanded to the structure
$(\N,X,g_1,\ldots,g_n,f)$ satisfying $f := \mathrm{op}(g_1,\ldots,g_n)$ and $f \in \FN$.

Although our algebras are formulated in a general way we
are mostly interested in subelementary classes of algebras characterizing some of the main computational complexity classes
(see \autoref{par:comp:classes} ).
It turns out that the oracles $X$ play important role in this and we use them as arguments (input) to the predicates of the complexity
classes. This is similar to the approach to computational complexity by finite models (see e.g. \cite{ebbinghaus}) where the
arguments are finite models.
Finite models contain interpretations of (finite) predicates which are comparable to our oracles.
We will thus restrict in our characterizations the
oracles to finite subsets of $\N$.

\PAR{Derivations}
We fix one class of algebras $\GA$ until the end of the paragraph.
\emph{Derivation terms} (or just a \emph{derivations})
are  the least set of symbols
containing the symbol $X_\ast$ and
the symbol $\mathrm{op}(d_1,\ldots,d_n)$
for each $n$-ary operator $\mathrm{op}$ of and derivations $d_1$, \ldots, $d_n$.
We fix the derivation terms into  the \emph{standard enumeration of derivations}:
$d_0, d_1, d_2 \ldots$ where the derivation
$\mathrm{op}(d_1,\ldots,d_n)$ is preceded by the symbols $d_1$,\ldots, $d_n$,
We identify the derivation terms with their indices in the standard
enumeration.
Thus for each algebra in the class we have a sequence $\ARR{f}$ enumerating its functions
such that for each $d \in \N$ the function $f_d$ has the derivation $d$.
Note that the the enumeration sequence is independent of the value of the oracle.

A typical use of enumerations $\ARR{f}$
will be in the construction of the first-order structures $(\N, X, \ARR{f})$ constituting
the standard models of the first-order theory called the \emph{recursive arithmetic of $\GA$} and designated by $\ARITHM 1{\GA}$.
This will be discussed in the following section.

\PAR{Clausal Definitions of Functions}\label{par:claus:def}
With the exception of the operator of bounded minimization (see \autoref{par:simple:operators}),
all function operators $f := \mathrm{op}(g_1,\ldots,g_n)$
discussed in this paper are specified by \emph{clausal definitions}.
A clausal definition
is obtained by a finite sequence $C_0$, \ldots, $C_k$ of finite sets of formulas
in $\LANG_2$. The formulas in $C_i$ are called \emph{clauses}.
The set $C_0$ consists of the single clause $\top \RW f(x) = y$. The set of clauses
$C_{i+1}$ is
obtained from the set $C_i$ by replacing one \emph{incomplete} clause $\varphi(x,\ARR{z}) \RW f(x) = y$ in $C_i$ which is
such that the formula $\varphi$ does not contain the variable $y$. If all clauses in $C_i$ are complete then $k = i$ and
the universal closure with the first-order quantifiers of the
formula $\bigwedge C_k$, abbreviated by $f := \mathrm{op}(g_1,\ldots,g_n)$,
is the \emph{clausal definition of $f$} (or the schema-of clausal definitions in the first order case).

The selected incomplete clause $\varphi(x,\ARR{z}) \RW f(x) = y$
in $C_i$ is \emph{refined} by choosing one of the numbered items in the following list. The set $C_{i+1}$ is then formed
to be like $C_i$ except
that the selected clause is replaced
by one or two clauses given in the chosen item:
\begin{enumerate}
\item
$\varphi(x,\ARR{z}) \land g(t(x,\ARR{z})) = v \RW f(x) = y$
where $g$ is one of $g_1$, \ldots, $g_n$, or $f$,
$t$ is a term of $\LANG_1$ in at most the indicated variables,
and $v$ is a \emph{new} variable, i.e. not occurring in $\varphi$ and different from $y$,
\item
$\varphi(x,\ARR{z}) \land v = 0 \RW f(x) = y$ and
$\varphi(x,\ARR{z}) \land v = S(w) \RW f(x) = y$ where the variable $v$ is one of $x,\ARR{z}$ and $w$ is new,
\item
$\varphi(x,\ARR{z}) \land v = 0 \RW f(x) = y$ and
$\varphi(x,\ARR{z}) \land v = (w_1,w_2) \RW f(x) = y$ where the variable $v$ is one of $x,\ARR{z}$ and $w_1$, $w_2$ are new,
\item
$\varphi(x,\ARR{z}) \land t_1(x,\ARR{z}) \mathrel{\mathrm{rel}} t_2(x,\ARR{z}) \RW f(x) = y$ and
$\varphi(x,\ARR{z}) \land  t_1(x,\ARR{z}) \not \mathrel{\mathrm{rel}} t_2(x,\ARR{z}) \RW f(x) = y$
where $t_1$ and $t_2$ are terms as above and $\mathrm{rel}$ is
either $=$ or $<$,
\item
$\varphi(x,\ARR{z}) \land t(x,\ARR{z}) = y \RW f(x) = y$ where $t$ is as above. Note that this clause is complete and cannot be
further refined.
\end{enumerate}
The clausal definition of $f$ is \emph{recursive} if the variable $f$ occurs in the antecedent of at least one clause, and
\emph{explicit} otherwise. A first-order variable other than $x$ and $y$
occurring in the antecedent of a clause is called a \emph{local} variable.

The three function operators  given in \autoref{par:opers} are not in the form of clausal definitions.
but it is straightforward to
bring them in into an equivalent clausal form in any theory extending $\IND \Delta_0^0$.
For instance, the operator of primitive recursion has an equivalent strict clausal form:
\begin{gather*}
 \bigl(\top \land x = 0 \land 0 = y \RW f(x) = y \bigr) \land {} \\
 \bigl (\top \land x = (v,p) \land v = 0 \land g(p) = z \land z = y \RW f(x) = y \bigr) \land {} \\
 \bigl (\top \land x = (v,p) \land v = S(w) \land f(w,p) = z \land h((w,z),p) = u \land u = y \RW f(x) = y \bigr) \ .
\end{gather*}
In the following we will not adhere to the strict form of clausal definitions if they can be equivalently rewritten
in an obvious way.

\PAR{Explicit Clausal Definitions}\label{par:claus:def:expl}
Provided that we already have the functions $g_1$, \ldots, $g_n$ defined, we wish to introduce the function $f$ specified by
an explicit clausal
definition $f := \mathrm{op}(g_1, \ldots, g_n)$ by a definitional extension of a theory
$T \vdash \IND\Delta_0^0{+}g_1,\ldots,g_n \in \FN$ in the language including $\LANG_1(g_1,\ldots,g_n)$
so that the extended theory $T_1$ proves $f := \mathrm{op}(g_1, \ldots, g_n)$ and $f \in \FN$.

This is easy to achieve for explicit clausal definitions which have the form
\[ \forall x,y,\ARR{w} \bigl(\bigwedge_i (\bigwedge_j \varphi_{i,j} \RW f(x) = y)\bigr)  \]
with the formulas $\varphi_{i,j}$ literals (atomic or their negations),
$\ARR{w}$ the local variables, and the variable $f$ not occurring in the antecedents of clauses.

The reader will note that the refinements of clauses are such that $T$ proves that for all $i \neq i'$ we have
\[ \forall x{,}y{,}\ARR{w}(\bigwedge_j \varphi_{i,j} \RW \lnot \bigwedge_j \varphi_{i',j})  \]
and
\[ \forall x \exists !y{,}\ARR{w} \bigvee_i \bigwedge_j \varphi_{i,j}  \ . \]
In other words, for each argument $x$ there is exactly one clause whose antecedent holds and its local variables plus $y$ are uniquely
determined.

The definitional extension of $T$ to $T_1$ is with the defining axiom:
\[ \forall v\bigl(v \in f \LRW \exists x{,}y{<}v\exists\ARR{w}(v = (x,y) \land \bigvee_i \bigwedge_j \varphi_{i,j})\bigr) \ . \]
$T_1$ then proves $f \in \FN$ and $f := \mathrm{op}(g_1, \ldots, g_n)$.

\PAR{Restrictions on Recursive Clausal Definitions}\label{par:rec:claus:def}
For a class of algebras $\GA$ with a
recursive function operator $f := \mathrm{op}(g_1, \ldots, g_n)$ we impose
two additional constraints on the form of its recursive clauses. They will allow to show that $f$ is primitive recursive in $\ARR{g}$.

Generally,
some \emph{measure} function $m$ must go down in recursive applications. This means that the above operator
is actually of the form $f := \mathrm{op}(g_1, \ldots, g_n,m)$ and in every model $(\N,X,g_1,\ldots,g_n,m)$
and for every recursive clause 
of the form $\varphi_1 \land f(t) = w \land \varphi_2 \RW f(x) = y$, the expanded
model must satisfy (the first-order universal closure of)
$\varphi \RW m(t) < m(x)$. The measure function for the operator of primitive recursion $f:= \mathrm{PR}(g,h)$
is the \emph{identity} function $I(x) = x$ and if this is the case
we do not explicitly include the measure as an argument to the operator because the above condition is simply
$\varphi \RW t < x$. As the first restriction on recursive definitions in our function
algebras we require that the measure function is the identity $I$.

If $n> 0$ then there is a second
restriction on recursive definitions that they must be in a \emph{parameterized} form
where their argument $x$ must be of the form $x = (v,p)$ with $p$ a parameter
shipped unchanged to all applications of $h \in \{ f, \ARR{g} \}$
in antecedents of clauses in the form $h(t(v,p,\ARR{z}),p) = w$
or $g_i(p) = w$. The requirement that $x = (v,p)$ is obviously required only in
clauses with applications of functions in  $\{ f, \ARR{g} \}$ in antecedents.
The requirement on the parameterization is an inessential restriction (the functions $g_i$ can always ignore the parameter $p$),
it permits the smooth transformation of first-order models
of theories for function algebras to second-order models
(see \autoref{cl:first:second:model:ara}).
The recursive operators used in this paper
(primitive recursion, bounded primitive recursion, doubly nested recursion) are all parameterized in this way.

\PAR{Reduction of Recursive Clausal Definitions to Primitive Recursion}  \label{par:rec:claus:red}
It should be clear that explicit clausal definitions plus $f := \mathrm{PR}(g,h)$
define all primitive recursive functions. We will now show the
converse. So we take an arbitrary recursive clausal definition of $f$ from the functions $g_1$, \ldots, $g_n$ with the measure $m$
whose set of clauses we designate by $C_0$. We will define $f$ by the operator of primitive recursion and by explicit definitions.

We could do it in way typically employed by logicians, namely by encoding the definition.
See, for instance, the treatment of nested ordinal recursion in Rose \cite{rose}. Computer scientists usually prefer
\emph{program transformations} over encoding where one definition is effectively translated into a simpler one.
In doing this, we will illustrate
the construction of
computer programs
directly in Peano Arithmetic. The main trick which makes this feasible is the
offsetting of the Cantor's
pairing function by one as reflected in its property $0 \neq (x,y)$.
This gives us a tool for the smooth development of programs directly in PA.
This is because
we obtain very simple codes of
finite sequences of natural numbers, called \emph{lists} in computer science.
There is, namely, a one-to-one correspondence
between lists and natural numbers because for every natural number $x$ there are unique numbers $n$, $x_1$, \ldots, $x_n$
such that $x = (x_1,\ldots,x_n,0)$. Thus $x$ can be taken as the code of the sequence $x_1$, \ldots, $x_n$.
The \emph{length} $L(x)$ of the list $x$ is $n$ and it satisfies the recursive clausal definition
$L(0) = 0 \land L(v,w) = L(w)+1$.

The clausal definitions are employed
in a slightly more refined form in our programming language CL (Clausal Language).
We have
been using the language (which comes with an integrated theorem prover for PA) in courses teaching computer programming and program
verification for the last twenty years \cite{voda:pa-and-cl}.

Returning to the clauses of the above recursive definition in the set $C_o$ we let $j_p$ to designate the number of
recursive applications in the $p$-th clause of $C_0$ (in some fixed ordering of $C_0$) and let $J := \max_p(j_p)$.
For all $p$ we assume w.l.o.g. that the successive recursive applications in the $p$-th clause of $C_0$ are numbered as
as
\[ f(t_{1}) = z_{1},  f(t_2) = z_{2}, \ldots, f(t_{j_p}) = z_{j_p}  \ . \]
Technically, we should have designated the terms by $t^{(p)}_i$ because they depend on $p$ but we will refrain from
doing so in order not to clutter the presentation.
Because the results of preceding recursions can be used in
succeeding ones, such recursion is called \emph{nested recursion}.

Just as it was demonstrated for the explicit clausal definitions in \autoref{par:claus:def:expl}
for each argument $x$ the antecedent
of exactly one clause in $C_0$ holds. For the demonstration just remove all recursive invocations of $f$ from the antecedents
of its clauses.
Denote the number of the clause applying to $x$ by $C(x)$.

We will translate the clauses in $C_0$ into an explicit clausal definition of an auxiliary
function $h$.
The function will be invoked in the form $h(x,c)$ where $c$ is a list such that if $p = C(x)$ then $i := L(c) \leq j_p$
and we have
$c = (z_1,\ldots,z_i,0)$ for some
$z_1$, \ldots, $z_i$ which are in that order the values of the first $i$ recursive calls to
$f$ in the $p$-th clause. If $i < j_p$
then the call $f_{i+1}(t_{i+1}) = z_{i+1}$ needs to be computed and this will be indicated by the function
$h$ yielding $(0,t_{i+1})$ with $0$ a \emph{tag} indicating this.
If $i = j_p$ then all recursive calls in the clause $p$ have been computed and
the value $y$ of $f(x)$ can be determined and $h(x,c)$ will yield $(1,y)$ with the tag $1$ indicating that the
value of $f(x)$ has been found.

We are now ready to describe the construction of the clauses for the function $h$.
This is done by successively
forming the sets of clauses $C_1$, $C_2$, \ldots $C_k$, \ldots
In forming the set $C_{k+1}$  we select a clause in the
set $C_k$ to which one of the following numbered transformations applies.
The set $C_{k+1}$ is obtained from $C_k$ by the replacement
of the selected clause by one or more clause specified in the applicable transformation step. If the selected clause is of the form:
\begin{enumerate}
\item
$\top \land \varphi \RW f(x) = y$ then the replacement clauses are
\begin{align*}
  \top \land  v = 0 \land  v  = w & {} \RW h(v) = w \\
  \top \land  v  = (x,c_0)  \land \varphi & {} \RW f(x) = y \ .
\end{align*}
\item
$\varphi_1 \land f(t_{i+1}) = z_{i+1} \land \varphi_2 \RW f(x) = y$ without $f$ occurring in $\varphi_1$ then the replacement clauses are
\begin{align*}
   \varphi_1 \land c_i = 0 \land (0,t_{i+1}) = w & {} \RW h(v) = w \\
   \varphi_1 \land c_i = (z_{i+1},c_{i+1}) \land \varphi_2 & {} \RW f(x) = y \ ,
\end{align*}
\item
$\varphi \land t = y \RW  f(x) = y$  without $f$ occurring in $\varphi$ then the replacement clause is
\begin{align*}
 \varphi \land  (1,t) = w & {} \RW h(v) = w \ .
\end{align*}
\end{enumerate}
The selection and replacement process will eventually terminate with the set $C_k$ where no clauses are selectable.
The clauses in $C_k$ then explicitly define $h$.
The reader will note that the clauses in $C_k$ have consequents of the form $h(v) = w$ instead of $h(x) = y$ but this is
inessential as the variables can be systematically renamed,
We also assume w.l.o.g. that the auxiliary variables
$v$, $w$, $c_0$, $\ldots$, $c_J$
introduced by the transformation are new.

The function $h$ is used in the following explicitly defined function $f_1$ which are easily transformable
into the strict
clausal form:
\begin{align*}
  &f_1( 0) = 0 {} \\
  &f_1(0,s_1) = (0,s_1) {} \\
   h(x,c) = 0 \RW {}&f_1((x,c),s_1) = ((x,c),s_1) \\
   h(x,c) = (0,z) \RW {}&f_1((x,c),s_1) = ((z,0),(x,c),s_1)  \tag{$\dagger_1$} \\
   h(x,c) = ((t_1,t_2),z) \land s_1 = 0 \RW {}&f_1((x,c),s_1) = ((x,c),s_1) \tag{$\dagger_2$} \\
   h(x,c) = ((t_1,t_2),z) \land s_1 = (0,s_2) \RW {}&f_1((x,c),s_1) = ((x,c),s_1) \\
    h(x,c) = ((t_1,t_2),z) \land s_1 = ((w,d),s_2)
       \RW {}&f_1((x,c),s_1) = ((w,d{\oplus} (z,0)),s_2)\ . \tag{$\dagger_3$}
\end{align*}
The real work is done in the marked clauses. The remaining ones are the \emph{default} clauses which make the function $f$
total although they cannot apply
when $f_1$ is correctly \emph{initialized} and used as $f_1^{\mu(x)}((x,0),0)$. This is a notation for the iteration of $f_1$
$\mu(x)$ times. The length of the iteration is given by the function $\mu(x)$ which will be
determined below.
The iteration function has the primitive recursive definition $f_1^0(s) = s$ and $f_1^{i+1}(s) = f_1\,f_1^i(s)$.
The function $\oplus$ used in the clause  $(\dagger_3)$ is the \emph{list concatenation} function
with the clausal definition:
\[   \bigl(0 \oplus y = y \bigr) \land \bigl((v,x)\oplus y = (v,x \oplus y)\bigr) \ . \]
The argument to the function $f_1(s)$ is a \emph{stack} $s$ which is a nonempty list of non-empty elements of the form $(x,c)$.
When $s = (x,c),s_1$ then the \emph{top} of the stack $(x,c)$ specifies that the $(L(c)+1)$-th recursive application
of $f$ in its $C(x)$-th clause should be computed by $h(x,c)$.
The clause $(\dagger_1)$ applies when there is such an application (because the tag yielded by $h$ is $0$)
and its argument is $z$. The stack is extended by \emph{pushing} $z$ on top of it together with the empty list $0$ signifying
that the first recursive application in the $C(z)$-th clause for $f$ should be computed (if there is such).

The clauses $(\dagger_2)$ and $(\dagger_3)$  apply if $h(x,c)$ yields $(1,z)$
signifying that the value of $f(x)$ has been computed to $z$. In the clause $(\dagger_2)$ the tail $s_1$ of the stack $s$ is
empty and we are essentially done. However, due to the fact that the length of iteration function $\mu(x)$ will give only an upper
bound, we yield the same stack by entering an \emph{idling loop}. When the iteration of $f_1$ eventually terminates
we will be able to read off the desired value of $f(x)$ from the stack as follows
$f(x) := T\,h\,H\,f_1^{\mu(x)}((x,0),0)$. The last identity is the desired definition of $f$ as primitive recursive in
the functions $\ARR{g}$, $m$.

The clause $(\dagger_3)$ applies when the stack $s_1$ is not empty and then $z$ is the value of the $(L(d)+1)$-th recursive
application of $f$ in the $C(w)$-th clause for $f$. The stack $s$ is \emph{popped} by removing $(x,c)$,
and the value $z$ extends the list $d$ before resuming the computation of the
$C(w)$-th clause.

It remains to find the function $\mu(x)$ giving the upper bound to the iterations of $f_1$.
The maximal length of the stack $s$ computing $f(x)$ is given by the measure $m:=m(x)$ but
during the computation of at most $J$ recursive applications in the antecedents of clauses for $f$
the stack will be repeatedly pushed and
popped. View the stack $s$ as coding the tail of the sequence:
\[ (a_0,c_0), (a_1,c_1), \ldots, (a_{m+1-L(s)},c_{m+1-L(s)}), \ldots, (a_{m},c_{m}) \]
which starts from the index $m+1-L(s)$ and with $a_0 = c_0 = a_1 = c_1 = \cdots = a_{m-L(s)} = c_{m-L(s)} = 0$.
With the stack initialized to $((x,0),0)$ we have $a_{m} = x$.
The function $m'(s) = \sum_{i \leq m}  L(c_i)\cdot J^i$ gives the weight of the stack $s$ during
the computation of $f(x)$.
We have $m'(s) < J^{m+1}$ and the reader can convince themselves that we have
$m'(s) < m'\,f_1(s)$
until the iteration of $f_1$ starts idling
by yielding the same stack. To bring the computation to idling it thus suffices to define $\mu(x) = J^{m(x)+1}$.

\section{Recursive Arithmetics}\label{sc:first:order:ra}

\PAR{Recursive Arithmetics}\label{par:algebra:theories:1}
Fix a class of function algebras $\GA$.
The class specifies
a first-order
theory  in the language $\LANG_1(X,\ARR{f})$,
designated by $\ARITHM1{\GA}$, and called the \emph{recursive arithmetic of~$\GA$}.
The theory consists of the basic axioms $\BASIC$, the \emph{oracle axioms}
$f_{X_\ast} := X$ and $f_{X_\ast} \in \FN$, and of the
\emph{operator axioms} $f_{\mathrm{op}(d_1,\ldots,d_n)} := \mathrm{op}(f_{d_1},\ldots,f_{d_n})$
and $f_{\mathrm{op}(d_1,\ldots,d_n)} \in \FN$
for each $n$-ary operator~$\mathrm{op}$ of~$\GA$
and each derivation $d_1$, \ldots, $d_n$.

Note that any structure $(\N,X)$ can be uniquely expanded
to the structure $(\N,X,\ARR{f})$ satisfying $\ARITHM1{\GA}$. The structures are called \emph{standard models of $\ARITHM1{\GA}$}.
In addition to the standard models, we also admit non-standard models
$(\STRUCT,X,\ARR{f})$ with $\STRUCT$ a model for~$\LANG_1$ and $X,\ARR{f} \subseteq \STRUCT$.

\PAR{Quasi-Terms and Quasi-Bounded Formulas}
We extend the notion of terms to \emph{quasi-terms} by allowing the expressions $f(t)$
in positions where a first-order term is permitted. They can be always unnested to the form
$\exists y(y = f(t) \land \varphi(y))$ or
$\forall y(y = f(t) \RW \varphi(y))$.
We abbreviate this to
$\exists y{=} f(t)\,\varphi(y)$ and $\forall y{=} f(t)\,\varphi(y)$ respectively, and call the quantifiers \emph{quasi-bounded}.
Bounded formulas extended with quasi-terms and quasi-bounded quantifiers are called \emph{quasi-bounded} formulas.

We note that by by unnesting the quasi-bounded terms in a quasi-bounded formula
we obtain a quasi-bounded formula and the ability to choose the kind of quasi-bounded quantifiers
makes the last formula equivalent to a $\Delta_1^0$ formula
over any theory proving $\COLL\Sigma_1^0$.

\PAR{Operators of
$\Delta_0$ Functions}\label{par:simple:operators}
We wish to connect the theory $\IND\Delta_0$ with a recursive arithmetic.
To that end we will need the following function operators
which are the first-order universal closures of the formulas named by the operators:
\begin{gather}
 f(x) = S(x)  \tag{$f:=\mathrm{S}$} \\
 f(x,y) = x+y   \land f(0) = 0  \tag{$f:=\oplus$} \\
 f(x,y) = x \cdot y \land f(0) = 0 \tag{$f:=\otimes$} \\
 (x < y \RW f(x,y) = 1)  \land  (x \ge y \RW f(x,y) = 0) \land f(0) = 0 \tag{$f:={<_\ast}$} \\
 f(x) = x           \tag{$f:=\mathrm{I}$} \\
 f(0,y,z) = y \land  f(S(x),y,z) = z \land f(0) = 0 \land f(x,0) = 0  \tag{$f:=\mathrm{D}$} \\
 g(x) = v \land h(x) = w \RW  f(x) = (v,w) \ .              \tag{$f:=\mathrm{P}(g,h)$}
\end{gather}
The nullary operators of \emph{successor}, \emph{addition}, \emph{multiplication}, \emph{identity}, (the characteristic function of)
\emph{order}, \emph{case-analysis}, and
the binary \emph{pairing} operator are not in a strict clausal form. However,
aAny weak theory proving $\BASIC$ permits to bring them into the
strict form of explicit clausal definitions.

In order to capture the properties of bounded quantifiers we introduce
the unary operator
of \emph{bounded minimization} $f := \mu(g)$ which is the first-order universal closure of:
\begin{align*}
  (&f(b,x) = z \RW z \leq b) \land {} \\
  (&f(b,x) = z \land y < b \land g(y,x) = 1  \RW  z \leq y) \land {} \\
  (&f(b,x) = z \land z < b \RW g(z,x) = 1) \land {} \\
   &f(0) = 0 \ .
\end{align*}
Informally, the function $f(b,x) := \mu_{z < b}[ g(z,x) = 1]$ yields the least $z < b$ satisfying $g(z,x)=1$ if there is such
and $b$ otherwise.
Bounded minimization can be brought (by a recursive search for $z$)
into an equivalent recursive clausal form, but this apparently
requires a theory stronger than $\IND\Delta_0^0$ (it is an open problem).
We can, however, extend by definition any theory $T \vdash \IND\Delta_0^0{+}g \in \FN$.
The defining axiom for $f$ is the closure by $\forall v$ of:
\begin{align*}
  v \in f \LRW v = (0,0) \lor \exists b,x,z {<} v \bigl(& v = ((b,x),z) \land ( z < b \land g(z,x) = 1 \lor z = b )
               \land {} \\
    & \forall y {<} z\, ((y,x),1) \notin g \bigr)
\end{align*}
whose RHS is  $\Delta^0_0$. By working in $T$ we prove
$f \in \FN$.
The proof of $0 \notin f$ and of the uniqueness property is straightforward.
The existence condition $\forall x' \exists y\, (x',y) \in f$ is trivial for $x' = 0$.
Otherwise, we have $x' = (b',x)$ for some
$b'$, $x$ and
we prove by $\Delta_0^0$ induction on $b$:
\[ b \leq b' \RW \exists z\leq b \bigl ( ( z < b \land  g(z,x) = 1 \lor z = b )\land
   \forall y {<} z\, ((y,x),1) \notin g  \bigr)\ .
\]
Now $f := \mu(g)$ easily follows.

We define the class $\DA$ of \emph{$\Delta_0^0$-algebras} to consists of the operators:
\[ \mathrm{S},{\oplus},{\otimes},{<_\ast},\mathrm{I},\mathrm{D},\mathrm{P},{\COMP},{\mu} \ .  \]
For any $X \subseteq \N$ the algebra  $\DA$ has the standard model $(\N,X,\ARR{f})$ with its functions polynomially bounded.

We henceforth require that the operators of every function algebra $\GA$
contains the operators of $\DA$.

The following \autoref{cl:char:form:1} asserts that the characteristic functions of quasi-bounded formulas of a recursive arithmetic
are denotations of its function constants. The lemma needs an auxiliary lemma about terms of such arithmetics.
The reader
will note that the proofs do not rely on any form of induction holding in the arithmetics.
That
the induction holds for all quasi-bounded formulas needs the two lemmas and it is asserted by \autoref{cl:induction:arithmetic:1}.

\begin{LEMMA}\label{cl:char:term:1}
For any recursive arithmetic $\ARITHM1{\GA}$,
any sequence of variables $\ARR{x} := x_0,\ldots,x_n$, and any quasi-term $t(\ARR{x})$
(whose variables are all indicated) there
is a derivation in $\GA$, designated (without the risk of confusion) by $t(\ARR{x})$, such that
\[ \ARITHM1{\GA} \vdash \forall \ARR{x}\,f_{t(\ARR{x})}(\ARR{x}) = t(\ARR{x}) \ . \]
\end{LEMMA}
\begin{proof}
By induction on the structure of the term $t$ while working in the recursive arithmetic $\ARITHM1{\GA}$.

If $t$ is $0$
we consider the following informal identity:
\[ f_{\mu(\mathrm{S} \COMP \otimes)}(b,x) = \mu_{y<b}[ (f_{\mathrm{S} \circ {\otimes}}(y,x) = 1] \ . \]
Thus we can prove $\forall x\, f_{\mu(\mathrm{S} \circ \otimes)}(x,x) = 0$ and
take the desired derivation $0(\ARR{x}) := \mathrm{Z} := \mu(\mathrm{S} \circ \otimes) \circ \mathrm{P}(\mathrm{I},\mathrm{I})$.

If $t$ is the variable $x_i$ ($i\leq n$) then we are looking for a derivation $d := x_i(\ARR{x}) $ s.t.
$\forall \ARR{x}\, f_d(\ARR{x}) = x_i$. Toward that end
we abbreviate $\mathrm{H} := \mathrm{D} \circ \mathrm{P}(\mathrm{Z},\mathrm{I})$.
It is easy to see that we have $\forall x\,f_{\mathrm{H}}(x) = H(x)$.
Likewise, $\mathrm{T} :=  \mathrm{D} \circ \mathrm{P}(\mathrm{S}\circ \mathrm{Z},\mathrm{I})$ is the derivation s.t.
$\forall x\,f_{\mathrm{T}}(x) = T(x)$.
We introduce the following abbreviations on derivations  $\mathrm{T}^0 := \mathrm{I}$ and
$\mathrm{T}^{i+1} := \mathrm{T} \circ \mathrm{T}^i$.
Now, if $i = n$ we set $d := \mathrm{T}^n$ and set $d := \mathrm{H} \circ \mathrm{T}^i$ otherwise.

If $t$ is one of
$S\,t_1(\ARR{x})$, $f_d\,t_1(\ARR{x})$, $t_1(\ARR{x})+t_2(\ARR{x})$, $t_1(\ARR{x})\cdot t_2(\ARR{x})$, or $(t_1(\ARR{x}),t_2(\ARR{x}))$
then we set the corresponding derivations $t(\ARR{x})$ to $\mathrm{S} \circ t_1(\ARR{x})$, $d \circ t_1(\ARR{x})$,
$\oplus \circ \mathrm{P}(t_1(\ARR{x}),t_2(\ARR{x}))$, $\otimes \circ \mathrm{P}(t_1(\ARR{x}),t_2(\ARR{x}))$,
or $\mathrm{P}(t_1(\ARR{x}),t_2(\ARR{x}))$ where the derivations $t_i(\ARR{x})$ are
obtained from the induction hypothesis.
\qed
\end{proof}

\begin{LEMMA}\label{cl:char:form:1}
For any recursive arithmetic $\ARITHM1{\GA}$,
any quasi-bounded formula $\varphi(\ARR{x})$ in $\LANG_1(X,\ARR{f})$
there is a derivation designated (without the risk of confusion) by $\varphi(\ARR{x})$ such that
\begin{gather*}
   \ARITHM1{\GA} \vdash \forall x\, f_{\varphi(\ARR{x})}(x) \leq 1 \\
  \ARITHM1{\GA} \vdash \forall \ARR{x} ( \varphi(\ARR{x}) \LRW f_{\varphi(\ARR{x})}(\ARR{x}) = 1 ) \ .
\end{gather*}
\end{LEMMA}
\begin{proof}
We do not give the proof that $f_{\varphi(\ARR{x})}$ is a $(0{-}1)$-valued function
because this will be obvious from the way the derivations are constructed. We work in $\ARITHM1{\GA}$ and proceed
by induction on the quasi-formulas $\varphi(\ARR{x})$
whose free variables are among
the indicated ones
and they are constructed from atomic formulas by negation, disjunction, bounded existential, and
quasi-bounded existential quantification.

If $\varphi$ is $t_1(\ARR{x}) < t_2(\ARR{x})$ we prove:
\[ \forall \ARR{x}\bigl(
 t_1(\ARR{x}) < t_2(\ARR{x}) \LRW f_{{<}_\ast}(f_{t_1(\ARR{x})}(\ARR{x}),f_{t_2(\ARR{x})}(\ARR{x})) = 1 \bigr)  \]
and it suffices to set the desired derivation $\varphi(\ARR{x}) := {<}_\ast \circ \mathrm{P}(t_1(\ARR{x}),t_2(\ARR{x}))$.

In the remaining cases we obtain the desired derivations from IH
and in a straightforward way from the properties proved below.

If $\varphi$ is $t_1(\ARR{x}) = t_2(\ARR{x})$ we note that $t_1 = t_2 \LRW t_1 \not< t_2 \land t_2 \not< t_1$ and prove
\[ \forall \ARR{x}\Bigl( t_1(\ARR{x}) {=} t_2(\ARR{x}) \LRW
  f_{\mathrm{D}}\bigl(f_{{<}_\ast}(f_{t_1}(\ARR{x}),f_{t_2}(\ARR{x})),f_{\mathrm{D}}(f_{{<}_\ast}(f_{t_2}(\ARR{x}),f_{t_1}(\ARR{x})),1,0),0
             \bigr) = 1  \Bigr) \ .\]

If $\varphi$ is $t(\ARR{x}) \in X$ we prove
$\forall \ARR{x}\bigl( t(\ARR{x}) \in X \LRW   f_{X_\ast}\,f_{t(\ARR{x})}(\ARR{x}) = 1 \bigr)$.

If $\varphi$ is $\lnot \psi(\ARR{x})$ we prove
$\forall \ARR{x}\bigl( \lnot \psi(\ARR{x}) \LRW  f_{\mathrm{D}}(f_{\psi(\ARR{x})}(\ARR{x}),1,0) = 1 \bigr)$.

If $\varphi$ is $\psi_1(\ARR{x}) \lor \psi_2(\ARR{x})$ we prove
$\forall \ARR{x}\bigl( \psi_1(\ARR{x}) \lor \psi_2(\ARR{x}) \LRW
   f_{\mathrm{D}}(f_{\psi_1(\ARR{x})}(\ARR{x}),f_{\psi_2(\ARR{x})}(\ARR{x}),1) = 1      \bigr)$.

If $\varphi$ is $\exists y {=} f_dt(\ARR{x})\,\psi(y,\ARR{x})$ we prove
\[  \forall \ARR{x}\bigl( \exists y {=} f_dt(\ARR{x})\,\psi(y,\ARR{x}) \LRW
          f_{\psi(y,\ARR{x})}(f_{t(\ARR{x})}(\ARR{x}),f_{\mathrm{I}}(\ARR{x}))=1\bigr) \ . \]
The final case is when $\varphi(\ARR{x})$ is
$\exists y {<} t(\ARR{x})\,\psi(y,\ARR{x})$.
We need an auxiliary function obtained by bounded minimization:
$f_{\mu(\psi(\ARR{x}))}(b,x) = \mu_{y < b}[ f_\psi(y,x) = 1]$. We then prove
\[ \forall \ARR{x}\Bigl(  \exists y < t(\ARR{x})\,\psi(y,\ARR{x})  \LRW
  f_{{<}_\ast}\bigl(f_{\mu(\psi(\ARR{x}))}(f_{t(\ARR{x})}(\ARR{x}),f_{\mathrm{I}}(\ARR{x})),f_{t(\ARR{x})}(\ARR{x}) \bigr)  = 1 \Bigr) \ . \]
\qed
\end{proof}

\begin{LEMMA}\label{cl:ara1:cl:explicit}
Every recursive arithmetic  $\ARITHM1{\GA}$  is closed under explicit clausal definitions, i.e.
for any explicit clausal definition
of $f$ from the function constants $f_{d_1}$, \ldots,  $f_{d_n}$
there is a function $f_d$ satisfying the clauses of the definition (after the replacement $f := f_d)$.
\end{LEMMA}
\begin{proof}
We take an explicit clausal definition as in the theorem. The clauses of the definition
are constructed as in \autoref{par:claus:def} into the set $C_k$ by refinements from the single clause
in $C_0$.
We first reason informally and
for each set $C_i$ $(i \leq k)$ we construct a set
$D_i$ of clauses with consequents of the form $f(x) = s$ for some quasi-terms $s$ composed
from $x$, $y$. local variables, and from the function constants of $\ARITHM1{\GA}$
where we abbreviate $g_1 := f_{d_1}$, \ldots,  $g_n :=f_{d_n}$.
During the process we fold the clauses into a single formula $f(x) = s$.

We work backwards from $k$ to $0$ and construct $D_k := C_k$
Assuming for $i < k$ that the set $D_{i+1}$ has been constructed, we construct the set $D_i$ according to the
form of the clause in $C_i$ refined into the set $C_{i+1}$.
If the clause was refined by the point 1) and
the clause corresponding to the refined clause in $D_{i+1}$ is
$\varphi(x,\ARR{z}) \land g(t(x,\ARR{z})) = v \RW f(x) = s(x,\ARR{z},v)$ with $g$ one of $g_1$, \ldots $g_n$, say $g_j$,
we place the clause
$\varphi(x,\ARR{z}) \RW f(x) = s(x,\ARR{z},g_{d_j}(t(x,\ARR{z})))$ into $D_i$. If the clause was
refined by 2) and the clauses corresponding to the refined clauses in $D_{i+1}$ are
$\varphi(x,\ARR{z}) \land v = 0 \RW f(x) = s_1(x,\ARR{z})$ and
$\varphi(x,\ARR{z}) \land v = S(w) \RW f(x) = s_2(x,\ARR{z},w)$ we place
$\varphi(x,\ARR{z}) \RW f(x) = f_D(v,s_1(x,\ARR{z}),s_2(x,\ARR{z},f_{\mathrm{Pr}}(v)))$
into $D_i$ where the function $f_{\mathrm{Pr}}$ is the predecessor function easily defined in the algebra by bounded
minimization. If the clause was
refined by 3) and the clauses corresponding to the refined clauses in $D_{i+1}$ are
$\varphi(x,\ARR{z}) \land v = 0 \RW f(x) = s_1(x,\ARR{z})$ and
$\varphi(x,\ARR{z}) \land v = (w_1,w_2) \RW f(x) = s_2(x,\ARR{z},w_1,w_2)$
we place
$\varphi(x,\ARR{z}) \RW f(x) = f_D(v,s_1(x,\ARR{z}),s_2(x,\ARR{z},f_{\mathrm{H}}(v).f_{\mathrm{T}}(v)))$ into $D_i$.
If the clause was
refined by 4) and the clauses corresponding to the refined clauses in $D_{i+1}$ are
$\varphi(x,\ARR{z}) \land t_1(x,\ARR{z}) \mathrel{\mathrm{rel}} t_2(x,\ARR{z}) \RW f(x) = s_1(x,\ARR{z})$ and
$\varphi(x,\ARR{z}) \land  t_1(x,\ARR{z}) \not \mathrel{\mathrm{rel}} t_2(x,\ARR{z}) \RW f(x) = s_2(x,\ARR{z})$
we place
\[\varphi(x,\ARR{z}) \RW f(x) = f_D(f_{\mathrm{rel}}(t_1(x,\ARR{z}),t_2(x,\ARR{z})), s_2(x,\ARR{z}),s_1(x,\ARR{z})) \]
into $D_i$.
Finally, if the clause was refined by 5) then
we place  $\varphi(x,\ARR{z}) \RW f(x) = t(x,\ARR{z})$ into $D_i$. The remaining unaffected clauses of $D_{i+1}$ are just copied to $D_i$.

At the end we have a single clause $\top \RW f(x) = s(x)$ in $D_0$ and we use \autoref{cl:char:term:1} to obtain the desired
derivation $d := s(x)$ for the function constant $f_d$.

To prove that the clauses for $f$ with $f$ replaced by $f_d$ in $D_i(f)$ are provable we work in
$\ARITHM1{\GA}$ and show by external induction that successively the clauses in $D_0(f_d)$, \ldots, $D_k(f_d)$
are all provable. The clauses $D_k(f_d)$ are the clauses of the explicit clausal definition.
\qed
\end{proof}

The next theorem asserts that the arithmetic $\ARITHM1{\DA}$ is essentially the
inductive theory $\IND\Delta_0^0(X)$:
\begin{THEOREM}\label{cl:induction:arithmetic:1}
$\ARITHM1{\DA}$ proves the induction principles $\IND[\Delta_0^0(X,\ARR{f})]$ and
that the functions $\ARR{f}$ are polynomially bounded.
Vice versa,
any inductive theory $\IND\Delta_0^0(X)$ can be extended by definitions to the theory
$\ARITHM1{\DA}$.
\end{THEOREM}
\begin{proof}
Take any formula $\varphi(b,\ARR{y})$ which is $\Delta_0^0$ in the language $\LANG(X,\ARR{f})$.
It suffices to prove the least number principle for $\varphi$ by
working in $\ARITHM1{\DA}$, So take any $\ARR{y}$ and assume $\varphi(b,\ARR{y})$ for some $b$.
By \autoref{cl:char:form:1}
there is a derivation $d := \varphi(b,\ARR{y})$
such that the theory proves that $f_d$ is the characteristic function of $\varphi$.
For $f_{\mu(d)}(b,x) = \mu_{y<b}[f_d(y,x) = 1]$ the theory proves that $f_{\mu(d)}(S(b),\ARR{y})$
is the least witness
for $\varphi(b,\ARR{y})$.

By external induction on derivations we prove that the functions $f_d$ are polynomially bounded.

The second part of the theorem directly follows from the way the operators of the algebra $\DA$ are defined. \qed
\end{proof}

\section{Provably Recursive Functions of $\ARITHM1{\GA}$}\label{sc:provably:rec}

\PAR{Provably Total Functions of~$\ARITHM1{\GA}$}
\label{par:provably-recursive}
A ($\Sigma_1^0$) \emph{provably total function} of an arithmetic\/~$\ARITHM1{\GA}$
is any function $f\colon \N \times 2^\N \to \N$
such that there is a $\Sigma_1^0$ formula without parameters~$\varphi(x,y) \in \LANG_1(X,\ARR{f})$
such that:
\begin{enumerate}[(a),leftmargin=*]
    \item $\ARITHM1{\GA} \vdash \forall x\exists! y\varphi(x,y)$,
    and
    \item for all $X \subseteq \N$ and all $x \in \N$
        the standard model $(\N,X,\ARR{f})$
        of $\ARITHM1{\GA}$
        satisfies $\varphi(x,f(x,X))$.
\end{enumerate}
We denote by $2^{{<}\N}$
the class of finite subsets of $\N$.
The restriction of a provably total function $f$ to the domain $\N \times 2^{{<}\N}$ is called a \emph{provably recursive}
function of $\ARITHM1{\GA}$. Clearly, $f$ is in general only recursive in $X$ but its restriction is recursive (with Ackermann's coding
of finite sets).

Provably total functions of $\ARITHM1{\GA}$
can be characterized using the ideas of Ferreira \cite{ferreira},
which are based on the following special form of Herbrand's theorem.
Its special case was proved by Krajíček, Pudlák, and Takeuti \cite{kpt}.


\begin{LEMMA}[Ferreira \cite{ferreira}]\label{cl:special-herbrand}
    Let~$T$ be a universal theory in a first-order language~$\LANG$.
    Suppose that
    $\exists\ARR{u}\,\forall\ARR{v}\,\varphi$
    is a consequence of\/~$T$,
    with $\varphi(\ARR{u},\ARR{v},x)$ an existential formula
    with only the indicated variables free.
    Then there are terms
    $\ARR{t}_1(x)$,
    $\ARR{t}_2(x,\ARR{v}_1)$, \ldots,
    $\ARR{t}_k(x,\ARR{v}_1,\ldots,\ARR{v}_{k-1})$
    of\/ $\LANG$ such that
    \begin{gather*}
       T \vdash
       \varphi(\ARR{t}_1(x),\ARR{v}_1,x) \lor
       \varphi(\ARR{t}_2(x,\ARR{v}_1),\ARR{v}_2,x) \lor \dotsb \lor
       \varphi(\ARR{t}_k(x,\ARR{v}_1,\ldots,\ARR{v}_{k-1}),\ARR{v}_k,x) \tag*{\qed}
       \text{.}
    \end{gather*}
\end{LEMMA}

%

\begin{THEOREM}\label{cl:prov:rec}
    The class of provably total functions of a recursive arithmetic $\ARITHM1{\GA}$
    is exactly the class of functions~$f$
    for which there is a derivation~$d$ in~$\GA$ such that
    all standard models $(\N,X,\ARR{f})$ of $\ARITHM1{\GA}$ satisfy
    $f_d(x) = f(x,X)$ for all $x \in \N$.
\end{THEOREM}
\begin{proof}
    Note that the function operators of $\GA$ are universal sentences.
    For the~$\supseteq$ inclusion of function classes
    from the claim of the theorem,
    take any function $f\colon \N\times 2^\N \to \N$ such that
    for some derivation~$d$
    for all $X \subseteq \N$ and all $x \in \N$
    we have $(\N,X,\ARR{f}) \models f_d(x) = f(x,X)$
    in the standard model of $\ARITHM1{\GA}$.
    The $\Sigma_1^0$ formula $\varphi(x,y)$ such that
    that $f$~is a provably total function of $\ARITHM1{\GA}$
    is simply $f_d(x) = y$.
    We have $\forall x\exists! y\varphi(x,y)$ in $\ARITHM1{\GA}$
    as $f_d \in \FN$, and the condition (b) is immediate.

    For the $\subseteq$ inclusion,
    take any function~$f\colon \N\times 2^\N \to \N$
    such that $f$~is a provably total function of $\ARITHM1{\GA}$.
    So there is
    a $\Sigma_1^0$ formula $\varphi(x,y)$ of $\LANG_1(X,\ARR{f})$
    satisfying conditions a) and~b) of
    \autoref{par:provably-recursive}.
    We will prove the theorem if we find a derivation~$d$ of $f$ s.t.
    $\ARITHM1{\GA} \vdash \forall x\varphi(x,f_d(x))$.

    Note that $\varphi(x,y)$ is $\exists z\varphi_0$
    for some $\Delta_0$ formula $\varphi_0(x,y,z)$.
    Hence, $\exists!y\varphi$ is equivalent in $\ARITHM1{\GA}$
    to $\exists! w\,\psi$ where $\psi(x,w)$ is the $\Delta_0$ formula
    $\exists y,z{<}w(w = (y,z) \land \varphi_0(x,y,z) \land
    \forall v{<}z\allowbreak\lnot\varphi_0(x,y,v))$.
    By \autoref{cl:char:form:1},
    there is a constant~$h := f_{\psi(x,w)}$
    which is the characteristic function of~$\psi(x,w)$.
    $\ARITHM1{\GA}$ thus proves $\exists!w\,h(x,w) = 1$.

    Although the $\BASIC$ part of $\ARITHM1{\GA}$
    contains some existential sentences,
    they can be eliminated at the expense of rewriting a few axioms
    using a newly introduced (proper) unary function symbol~$\mathit{Pr}$:
    \begin{gather}
        0 < z \RW (H(z),T(z)) = z
                  \land H(z) < z \land T(z) < z
            \tag{$\mathrm{P2'}$}\\
        \mathit{Pr}(0) = 0 \land
            \bigl( 0 < x \RW
                   S(\mathit{Pr}(x)) = x \land \mathit{Pr}(x) < x
            \bigr)\text{\rlap{.}}
            \tag{$\mathrm{N10'}$}
    \end{gather}
    The resulting theory $\ARITHM1{\GA}'$
    is a conservative extension of $\ARITHM1{\GA}$,
    equivalent to the extension of $\ARITHM1{\GA}$
    by definition
    $\mathit{Pr}(x) = y \LRW (x = 0 \RW y = 0) \land (x > 0 \RW x = S(y))$
    of~$\mathit{Pr}$.
    In particular,
    $\ARITHM1{\GA}'$ proves $\exists!w\,h(x,w) = 1$.

    By compactness,
    a finite subset $T_1 \uplus T_2$ of $\ARITHM1{\GA}'$ proves
    $\exists w\,h(x,w) = 1$.
    Here, sentences~$T_1$ are universal, i.e.,
    some of $\BASIC'$ axioms, instances of operators of $\GA$,
    and the uniqueness parts $\forall x\forall y_1\allowbreak\forall y_2
    (f_{d_i}(x)=y_1\land f_{d_i}(x)=y_2 \RW y_1 = y_2 )$
    of axioms $f_{d_i} \in \FN$
    for some derivations $d_1$, \ldots,~$d_n$.
    Sentences~$T_2$ are~$\forall\exists$:
    the existence parts $\forall x\exists w\,f_{d_i}(x) = w$
    of $f_{d_i} \in \FN$.
    Thus,
    $T_1 \vdash \exists w\,\exists\ARR{u}\,\forall\ARR{v}\,(
        \bigwedge_{i=1}^n f_{d_i}(u_i) = v_i \RW h(x,w) = 1)$.
    Let us abbreviate the antecedent to $\ARR{g}(\ARR{u}) \wedgeq \ARR{v}$.

    By \autoref{cl:special-herbrand},
    we have $n$-tuples of terms
    $\ARR{s}_1(x)$, \ldots, $\ARR{s}_k(x,\ARR{u}_1,\ldots,\ARR{u}_{k-1})$
    and terms $t_1(x)$, \ldots, $t_k(x,\ARR{u}_1,\ldots,\ARR{u}_{k-1})$
    such that
    $T_1$ (and hence also $\ARITHM1{\GA}'$)
    has as its consequence the disjunction of
    \begin{align*}
        \ARR{g}(\ARR{s}_1(x)) = \ARR{v}_1 &\RW h(x,t_1(x)) = 1 \\
        \ARR{g}(\ARR{s}_2(x,\ARR{v}_1)) = \ARR{v}_2 &\RW h(x,t_2(x,\ARR{v}_1)) = 1 \\
        &\cdots \\
        \ARR{g}(\ARR{s}_k(x,\ARR{v}_1,\ldots,\ARR{v}_{k-1})) = \ARR{v}_k &\RW
         h(x,t_k(x,\ARR{v}_1,\ldots,\ARR{v}_{k-1})) = 1
    \end{align*}
    which can be rewritten as the formula
    \begin{equation}
        \begin{split}
            &\ARR{g}(\ARR{s}_1(x)) = \ARR{v}_1
             \land \ARR{g}(\ARR{s}_2(x,\ARR{v}_1)) = \ARR{v}_2 \land \dotsb
             \land \ARR{g}(\ARR{s}_k(x,\ARR{v}_1,\ldots,\ARR{v}_{k-1}))
                      = \ARR{v}_k \RW {}\\
            &\qquad h(x,t_1(x)) = 1 \lor h(x,t_2(x,\ARR{v}_1)) = 1 \lor
                 \dotsb
                 \lor h(x,t_k(x,\ARR{v}_1,\ldots,\ARR{v}_{k-1})) = 1
            \text{\rlap{.}}
        \end{split}
        \label{eq:bigclause}\tag{$\theta_1$}
    \end{equation}
    Thus $\ARITHM1{\GA}' \vdash
    \forall\ARR{v}_1\dotso\forall\ARR{v}_k\eqref{eq:bigclause}$.
    The last formula can be abbreviated
    as the quasi-bounded formula
    \begin{equation}
        h(x,q_1(x)) = 1 \lor h(x,q_2(x)) = 1 \lor
        \dotsb \lor h(x,q_k(x)) = 1
        \label{eq:quasibigclause}\tag{$\theta_2$}
    \end{equation}
    where the quasi-terms $q_i(x)$ are obtained
    by substituting the respective left-hand sides
    from the antecedent of~\eqref{eq:bigclause}
    for the variables $\ARR{v}_1$, \ldots, $\ARR{v}_k$,
    i.e., $q_1(x)$ is $t_1(x)$, $q_2(x)$ is
    $t_2\bigl(x,\allowbreak g_1(s_{1,1}(x)),\dotsc,g_n(s_{1,n}(x))\bigr)$,
    etc.

    Note that \autoref{cl:char:term:1} can be proved also
    for $\ARITHM1{\GA}'$ in $\LANG_1(\mathit{Pr},X,\ARR{f})$.
    In particular, any quasi-term of the form $\mathit{Pr}(t(x))$
    is computed by the bounded minimization
    $\mu_{z<t(x)}[t(x) < S(S(z))]$
    with the derivation
    $\mathrm{Pr} :=
      \mu({<_*} \mathbin{\circ}
          \mathrm{P}(\mathrm{T},
                     \mathrm{S} \circ\mathrm{S}\circ\mathrm{H}))
      \mathbin{\circ} \mathrm{P}(\mathrm{I},\mathrm{I})
      \mathbin{\circ} t(x)$.
    We can thus replace quasi-terms $q_i(x)$ in~\eqref{eq:quasibigclause}
    with applications of the respective functions~$f_{q_i(x)}$,
    thus obtaining an equivalent quasi-bounded formula
    \[
        h(x,f_{q_1(x)}(x)) = 1 \lor \dotsb \lor h(x,f_{q_k(x)}(x)) = 1\text{\rlap{.}}
        \tag{$\theta_3$}\label{eq:reducedclause}
    \]
    As this is in the original
    language~$\LANG_1(X,\ARR{f})$,
    we have $\ARITHM1{\GA} \vdash \eqref{eq:reducedclause}$ by conservativity.

    Since for each~$x$ a unique~$w$ satisfies $\psi(x,w)$
    and this is one of $f_{q_i(x)}(x)$,
    we can obtain it by simply testing the values~$f_{q_i(x)}(x)$
    one after another, informally:
    \[\begin{array}{llllll}
        f_{\mathrm{D}}\bigl(h(x,f_{q_1(x)}(x)),\;
            & f_{\mathrm{D}}\bigl(h(x,f_{q_2(x)}(x)),\;
                & f_{\mathrm{D}}\bigl(\cdots,\;
                    f_{\mathrm{D}}\bigl(h(x,f_{q_{k-1}(x)}(x)),\;
                                            & f_{q_{k}(x)}(x),\\
            &   &                           & f_{q_{k-1}(x)}(x)\bigr),
                                            \ldots\bigr),\\
            &   & f_{q_2(x)}(x)\bigr),\\
            & f_{q_1(x)}(x)\bigr)\,\text{.}
    \end{array}\]
    Since a derivation~$e$ of the above function exists in $\GA$,
    so does $d := \mathrm{H}\circ e$ which is such that
    $\ARITHM1{\GA} \vdash
    \forall x\varphi(x,f_d(x))$,
    thus completing the proof.
    \qed
\end{proof}

\section{Second-Order Recursive Arithmetics}
\label{sec:ara2}
In the second draft we will modify the presentation in this section
in order to be able to accommodate the theories $\ARITHM2{\GA}{+}\WKL$. For the time being we have a problem
with the formulation of correct assumptions for the
\autoref{cl:first:second:model:ara}.
The proofs of the remaining theorems will be simplified once we fix this problem. Also \autoref{cl:first:second:conservative}
shuld be generalized to all function subalgebras of PRA.

\PAR{Second-Order Theories for Function Algebras}\label{par:algebra:theories:new}
Fix a class of function algebras $\GA$. The algebras determine a second-order theory in the language $\LANG_2$
designated by $\ARITHM2{\GA}$ and called the \emph{second-order recursive arithmetic of $\GA$}.
The theory is axiomatized by $\BASIC$ plus the following set existence axioms:
\begin{gather}
 \forall g_1,\ldots,g_n {\in} \FN\exists f {\in} \FN\, f:= \mathrm{op}(g_1,\ldots,g_n), \tag{$\mathrm{op}$} \label{eq:op2}
\end{gather}
one for each operator $f := \mathrm{op}(g_1,\ldots,g_n)$ of $\GA$.
There is the \emph{oracle} axiom:
\[
 \forall X \exists f{\in}\FN\, f := X_\ast \tag{$\mathrm{X}_\ast$}
 \label{eq:oracle2}
\]
and the
\emph{function comprehension} axiom $\mathrm{FC}$:
\[
 \forall p \forall f {\in} \FN \exists X\,X := \mathrm{FC}(f,p)  \tag{$\mathrm{FC}$}
 \label{eq:fc}
\]
where $X := \mathrm{FC}(f,p)$ abbreviates $\forall x(x \in X \LRW f(x,p) = 1)$.

\PAR{Lifting of First-Order Sentences to Second-Order}\label{par:lifing}
The language of second-order arithmetic $\LANG_2$ is in general not an extension
of the first-order language $\LANG_1(X,\ARR{f})$.
In order to characterize the relationship of the second-order arithmetic
$\ARITHM2{\GA}$ with its first-order counterpart $\ARITHM1{\GA}$
we will employ a syntactic transformation called lifting.

Fix a language~$\LANG_1(X,\ARR{f})$.
For $k \in \N$,
let $\ARR{f}_k$ be the initial part $f_0,\ldots,f_{k-1}$ of~$\ARR{f}$,
and let $\Phi_k$ be the set of \emph{definitions} of symbols $f \in \ARR{f}_k$,
i.e., either the oracle axiom $f := X$
or the operator axiom $f := \mathrm{op}(\ARR{g})$ of $\ARITHM1{\GA}$
respective to the constant~$f$.
Note that $\Phi_0$~is empty,
and if there is $f_d := \mathrm{op}(f_{d_1},\ldots,f_{d_n})$ in~$\Phi_k$,
then there are also the definitions of $f_{d_1}$, \ldots, $f_{d_n}$ in~$\Phi_k$.

Let $\varphi$ be any $\LANG_1(X,\ARR{f})$ sentence.
We define $\Phi_\varphi \coloneq \Phi_{k}$ and  $\ARR{f}_\varphi \coloneq \ARR{f}_{k}$
where $k = d+1$ for the highest~$d$ such that~$f_d$ occurs in~$\varphi$,
and $k = 0$ if no such~$d$ exists.
If we treat set constants $X$ and $\ARR{f}_{\varphi}$
as set variables of~$\LANG_2$, then
$\forall X\,\forall\ARR{f}_\varphi((\bigwedge \Phi_\varphi) \RW \varphi)$
is a formula of~$\LANG_2$. We write it as $\varphi{\uparrow}$,
and call it the \emph{lifted form} of~$\varphi$.

\begin{LEMMA}\label{cl:second:exists:Phi_d}
$\ARITHM2{\GA} \vdash \forall X\,\exists\ARR{f}_k\bigwedge\Phi_{k}$
\end{LEMMA}

\begin{proof}
    The claim is easily proved by external induction on~$k$.
    The inductively constructed proofs in $\ARITHM2{\GA}$
    use its axioms \eqref{eq:oracle2} and~\eqref{eq:op2}.
    \qed
\end{proof}
%

\PAR{Relationships of First-Order and Second-Order Arithmetics}
Lifting enables us to express the connection of
a first-order and the respective second-order recursive arithmetics
through the following
analogues of standard notions of extensions and conservativity.

We say that a second-order theory~$T$
is a \emph{lifted extension} of $\ARITHM1{\GA}$
if it proves the lifted forms of all its theorems, i.e.,
for any sentence $\varphi \in \LANG_1(X,\ARR{f})$
such that $\ARITHM1{\GA} \vdash \varphi$
we have $T \vdash \varphi{\uparrow}$.

For a class of sentences $\varGamma \subseteq \LANG_1(X,\ARR{f})$,
a second-order theory~$T$ is
\emph{lifted $\varGamma$~conservative} over $\ARITHM1{\GA}$
if all $\varGamma$ sentences
whose lifted forms are theorems of~$T$
are theorems of $\ARITHM1{\GA}$, i.e.,
for any sentence $\varphi \in \varGamma$
such that $T \vdash \varphi{\uparrow}$
we have $\ARITHM1{\GA} \vdash \varphi$.

The following \autoref{cl:first:second:extension} straightforwardly shows
that $\ARITHM2{\GA}$ is a lifted extension of $\ARITHM1{\GA}$.
Conservativity is more involved:
We first show how certain models of first-order arithmetics can be extended
to second-order models in \autoref{cl:first:second:model:ara}.
We then show $\Pi^0_2$~conservativity
for the special case of polynomially bounded arithmetics in \autoref{cl:first:second:conservative}.

\begin{THEOREM}\label{cl:first:second:extension}
$\ARITHM2{\GA}$ is a lifted extension of $\ARITHM1{\GA}$.
\end{THEOREM}
\begin{proof}
    Fix a class of function algebras~$\GA$.
    Take any sentence~$\varphi$
    such that $\ARITHM1{\GA} \vdash \varphi$,
    and any second-order structure
    $(\STRUCT,\SETS) \models \ARITHM2{\GA}$.
    Consider $\varphi{\uparrow}$
    and~the respective $\Phi_\varphi = \Phi_k$ for some~$k$,
    which has $X$ and~$\smash{\ARR{f}_k}$ as its free set variables.
    Choose any sets $Y$, $\smash{\ARR{g}_k} \in \SETS$ so that
    $(\STRUCT,\SETS) \models \Phi_\varphi(Y,\smash{\ARR{g}_k})$.

    Since $(\STRUCT,\SETS)$ satisfies \eqref{eq:oracle2}
    and \eqref{eq:op2}~axioms for operators of~$\GA$,
    $\smash{\ARR{g}_k}$ can be extend by external induction
    to~$\smash{\ARR{g}_d}$
    such that $(\STRUCT,\SETS) \models \Phi_d(Y,\smash{\ARR{g}_d})$
    for all $d \in \N$.
    Thus $(\STRUCT,Y,\ARR{g})$ for $\ARR{g} = \bigcup_{d\in\N}\ARR{g}_d$
    is a model of~$\ARITHM1{\GA}$,
    and hence a model of~$\varphi$ as well.
    Then, however also $(\STRUCT,\SETS) \models \varphi(Y,\ARR{g}_k)$.
    \qed
\end{proof}

\begin{THEOREM}\label{cl:second:rca}
$\ARITHM2{\GA} \vdash \RCA^-$
\end{THEOREM}

\begin{proof}
    The two theories share the $\BASIC$ axioms.
    Since $\GA$ includes~$\circ$ along with other operators of $\DA$,
    we have $\ARITHM2{\GA} \vdash \eqref{eq:composition2}$.
    It thus remains to prove $\COMPR[\Delta^0_0]$ and~$\INDAX$
    in $\ARITHM2{\GA}$.

    For $\COMPR[\Delta^0_0]$,
    take a~$\Delta^0_0$ formula~$\varphi(x,\ARR{y},\ARR{X})$
    with all parameters
    among those in non-empty sequences $\ARR{y}$ or~$\ARR{X}$.
    Let $\varphi'(x,\ARR{y},X)$ be the result
    of replacing every occurrence of~$\tau \in X_i$
    in $\varphi(x,\ARR{y},\ARR{X})$
    with $(S^i(0),\tau) \in X$.
    Considering~$X$ as the oracle set constant,
    obtain the characteristic function
    $f' \in \ARR{f}$ of~$\varphi'(x,\ARR{y})$
    from \autoref{cl:char:form:1}.
    Note that $\ARITHM1{\GA} \vdash
        \forall x\forall\ARR{y}\,f'(x,\ARR{y}) \leq 1 \land
        \forall x\forall\ARR{y}(f'(x,\ARR{y}) = 1 \LRW
                                 \varphi'(x,\ARR{y}))$
    and denote this formula by~$\psi$.
    $\ARITHM2{\GA}$ proves
    $\forall X\,\forall\ARR{f}_\psi(\Phi_\psi \RW \psi)$
    by the extension theorem~\ref{cl:first:second:extension}
    and $\forall X\,\exists\ARR{f}_\psi\,\Phi_\psi$
    by \autoref{cl:second:exists:Phi_d}.

    Let us now reason in $\ARITHM2{\GA}$:
    Take any numbers~$\ARR{y}$ and any sets~$\ARR{X}$,
    and obtain for every $X_i \in \ARR{X}$
    its characteristic function $g_i$ by~\eqref{eq:oracle2}~axiom.
    By applying the respective \eqref{eq:op2}~axioms
    for $\DA$~operators,
    obtain the function~$g$
    such that $\forall x\forall p\bigl(g(x,p) = 1
        \LRW \bigvee_i(H(x) = S^i(0) \land g\,T(x) = 1)\bigr)$
    (cf.\ppsp Lemmas \ref{cl:char:form:1} and~\ref{cl:char:term:1}).
    Obtain~$Y$ from~$g$ using the axiom~\eqref{eq:fc} with $p = 0$.
    Notice that $\forall x((S^i(0),x) \in Y \LRW x \in X_i)$.
    Since $\forall X\,\exists\ARR{f}_\psi\,\Phi_\psi$,
    take some~$\ARR{f}_\psi$ for which~$\Phi_\psi(Y,\ARR{f}_\psi)$ holds.
    We then have~$\psi(Y,\ARR{f}_\psi)$,
    and $f'$ among~$\ARR{f}_\psi$
    is such that $f'(x,\ARR{y}) = 1 \LRW \varphi'(x,\ARR{y},Y)
        \LRW \varphi(x, \ARR{y}, \ARR{X})$
    for all~$x$ and~$\ARR{y}$.
    Hence the \eqref{eq:fc}~axiom gives us for $f'$ and $p = (y_1,\ldots,y_n)$
    a~set~$X$ such that $\forall x(x \in X \LRW
         \varphi(x,\ARR{y},\ARR{X}))$.
    This concludes the proof of~$\COMPR[\Delta^0_0]$.

    Let us prove $\INDAX$ in $\ARITHM2{\GA}$:
    Take any~$X$ and assume $0 \in X$, and
    $\forall x(x \in X \RW S(x) \in X)$.
    Suppose there is some $z \notin X$.
    Obtain the set $X'$ such that $\forall x(x \in X' \LRW H(x) \notin X)$
    from~$\COMPR[\Delta^0_0]$,
    its characteristic function~$g'(x) = 1 \LRW x \in X'$
    from~\eqref{eq:oracle2},
    and its bounded minimization $f'(b,p) = \mu_{y<b}[g'(y,p) = 1] =
    \mu_{y<b}[y \notin X]$
    from the respective \eqref{eq:op2}~axiom.
    Now $y = f'(z,0)$ is the least number such that $y \notin X$.
    Since $0 \in X$, there must be~$u$ such that $y = S(u)$.
    By minimality of~$y$, we have $u \in X$, but then $S(u) = y \notin X$
    contradicts the second assumption of the induction axiom.
    Thus $\forall x\,x\in X$.
    \qed
\end{proof}

\begin{COR}\label{cl:simple:rca}
The theories $\RCA^-$ and $\ARITHM2{\DA}$ are equivalent.
\end{COR}
\begin{proof}
    $\ARITHM2{\DA} \vdash \RCA^- $
    is a special case of \autoref{cl:second:rca}.
    For $\RCA^- \vdash \ARITHM2{\DA}$,
    $\RCA^-$ includes $\circ$,
    and we can easily prove \eqref{eq:oracle2}, \eqref{eq:fc},
    and \eqref{eq:op2} for operators $\mathrm{S}$, $\oplus$,
    $\otimes$, ${<_\ast}$, $\mathrm{I}$, $\mathrm{D}$,
    $\mathrm{P}$ and~$\mu$
    within $\RCA^-$ using $\COMPR[\Delta^0_0]$.
    For instance, in the latter case, we obtain~$f := \mu(g)$ as
    \begin{align*}
        w \in f &\LRW  w = (0,0) \lor
            \exists b{\leq}w\,\exists x{\leq} w\,\exists z{\leq}b
            \bigl(w = ((b,x),z) \land{}\\
                &\bigl( ((z,x),1) \in g \land
                        \forall y{<}z\,((y,x),1) \notin g \lor
               z = b \land \forall y{\leq}b\,((y,x),1) \notin g
                \bigr)
            \bigr).
    \end{align*}
    The function value uniqueness part of $f \in \FN$
    is immediate. The function value existence part follows from $\INDAX$
    for the set~$X$ such that
    $v \in X \LRW \forall
        w{\leq}v\,\exists x{,}b{\leq}w\,\exists z{\leq}b\,
                (w = (b,x) \land (w,z) \in f)$
    obtained by $\COMPR[\Delta^0_0]$.
    \qed
\end{proof}

\newcommand{\TEMP}
{
\begin{LEMMA}\label{cl:first:second:model:ara}
If $(\STRUCT, X, \ARR{f}) \models \ARITHM1{\GA}$ and
$\STRUCT[I]$ is a proper initial segment of $\STRUCT$ such that
$(\STRUCT[I], X \cap \STRUCT[I],\{f_d \cap \STRUCT[I]\}_{d \in \N}) \models \ARITHM1{\GA}$
then for
\[ \SETS := \mathrm{FC}_{\STRUCT[I]}(\STRUCT,X,\ARR{f}) :=
    \{\, \{\, x {\in} \STRUCT[I] \mid (\STRUCT,X,\ARR{f}) \models g(x,p) = 1 \} \mid
         g \in \ARR{f}{,}\ p \in \STRUCT[I] \,\} \]
we have
$(\STRUCT[I],\SETS) \models \ARITHM2{\GA}$
and
$X \cap \STRUCT[I] \in \SETS$ as well as $f \cap \STRUCT[I] \in \FN^{(\STRUCT[I],\SETS)}$ for all $f \in \ARR{f}$.
\end{LEMMA}
\begin{proof}
    Let ${\GA}$, $\STRUCT[N] := (\STRUCT,X,\ARR{f})$, $\STRUCT[I]$, and  $\SETS$
    be as in the assumptions of the lemma.
    Abbreviate $\FN := \FN^{(\STRUCT[I],\SETS)}$ and 
$\STRUCT[N]\cap \STRUCT[I] := (\STRUCT[I], X \cap \STRUCT[I],\{f_d \cap \STRUCT[I]\}_{d \in \N})$.

For every function $f'' \in \ARR{f}$ and a parameter $p \in \STRUCT[I]$ the set
$\{v {\in} \STRUCT[I] \mid \STRUCT[N]\models f''(v,p) = 1 \}$ is in $\SETS$.
For the duration of this proof we say that the set \emph{is obtained from $f''$ and $p$ by function comprehension}. 
We obtain $h_1, f' \in \ARR{f}$ such that
\begin{align*}
& \STRUCT[N] \models 
  \forall y{,}x{,}q\ h_1(y,x,q) \stackrel{\text{\ref{cl:char:term:1}}}{=} f''((x,y),q) \land {} \\
& \STRUCT[N] \models \forall x{,}c{,}q\ f'(x,c,q) \stackrel{\text{\ref{cl:char:term:1}}}{=}  f_{\mu(h_1)}(c,x,q) \ . \\
\end{align*}
It should be obvious that for all $x,c,q \in \STRUCT$ we have
\begin{align*}
 \STRUCT[N] \models f'(x,c,q) = y \LRW  \forall y' {<} y\ f''((x,y'),p) \neq 1 \land (
                                         y < c \land  f''((x,y),q) \neq 1 \lor y = c) \ .
\end{align*}
For greater intuitiveness we might abbreviate the defining formula for $f'$ by $f'(x,c,q) = \mu_{y {<} c}[ f''((x,y),q) = 1]$
and call the function $f'$ \emph{associated to $f''$}.
For $p \in \STRUCT[I]$ and $c \in \STRUCT \setminus \STRUCT[I]$ define 
the 
set $\mathrm{Par}(f,c,p) := \{ (x,y)^{\STRUCT} \mid x \in \STRUCT[I], \STRUCT[N] \models f'(x,c,p) = y\}$, which is obviously such that
$f \subseteq \mathrm{Par}(f,c,p) \subseteq \STRUCT$.

We have the following characterization of the class $\FN$.
\begin{quote}
If $f \in \SETS$ is obtained from $f''$ and $p$ by function comprehension and $f'$ is associated to $f''$  then 
$f \in \FN$ iff for all $c \in \STRUCT \setminus \STRUCT[I]$ we have $\mathrm{Par}(f,c,p) \subseteq f$.
\end{quote}
\UNFIN

After these preliminary definitions we start with the proof the lemma.
We have $X \cap \STRUCT[I] \in \SETS$ because by \autoref{cl:char:term:1} there is a $g \in \ARR{f}$ s.t.
$\STRUCT[N] \models \forall x\forall p\,g(x,p) = f_{X_\ast}(x)$ and so
$X  \cap \STRUCT[I] = \{ x \in \STRUCT[I] \mid \STRUCT[N] \models g(x,0) = 1 \} \in \SETS$.

For any $d \in \N$ we have $f_d \in \FN^{\STRUCT[N]}$ and we wish to show that $f_d \cap \STRUCT[I] \in \FN$.
From the assumptions of the lemma we have $\STRUCT[N] \cap\STRUCT[I] \models f_d \in \FN$.
Thus it suffices to show $f_d \cap \STRUCT[I] \in \SETS$ which follows from
$\STRUCT[N] \models \forall v{,}p(g(v,p) = 1 \LRW v \in f)$
for a $g \in \ARR{f}$ obtained by \autoref{cl:char:form:1} and it suffices to take $p := 0$.

It remains to show that $(\STRUCT[I],\SETS) \models \ARITHM2{\GA}$. The axioms $\BASIC$ are trivially satisfied.

For the proof of $(\STRUCT[I],\SETS) \models \mathrm{X}_\ast$ take any $Z \in \SETS$. So there is
a $g \in \ARR{f}$ and $p \in \STRUCT[I]$ s.t. $Z = \{ x \in \STRUCT[I] \mid \STRUCT[N] \models g(x,p) = 1 \}$. We can assume
w.l.o.g. that
$g$ is a $(0{-}1)$-valued function because if not, we define by explicit clausal definition:
\[ g'(0) = 0 \land ( g(x,q) = 1 \RW  g'(x,q) = 1) \land ( g(x,q) \neq 1 \RW  g'(x,q) = 0)  \]
and use $g'$ which is in $\ARR{f}$ by \autoref{cl:ara1:cl:explicit}.
\UNFIN We need to define 
$f := \mathrm{Par}(g,p) \in \FN$. It is the characteristic
function of $Z$ because  $(\STRUCT[I],\SETS) \models f := \mathrm Z_\ast$.
This will be used also in the contract below.

For the proof of $(\STRUCT[I],\SETS) \models \mathrm{FC}$ we take any $p \in \STRUCT[I]$ and $f \in \FN$.
We have $f \in \SETS$ and so $f = \{ v \in \STRUCT[I] \mid g(v,q) = 1 \}$ for some $g \in \ARR{f}$ and $q \in \STRUCT[I]$.
By \autoref{cl:char:term:1} obtain a $g' \in \ARR{f}$ s.t. $\STRUCT[N] \models \forall x{,}p{,}q\ g'(x,p,q) = g(((x,p),1),q)$
and take the set $Z := \{ x \in \STRUCT[I] \mid \STRUCT[N] \models g'(x,p,q) = 1 \} \in \SETS$. 
The restrictions of $g'$ and $g$ to
$\STRUCT[I]$ are in $\SETS$ by the above and so 
for all $x \in \STRUCT[I]$ we have
\[(\STRUCT[I],\SETS) \models  x \in Z \LRW (g' \cap \STRUCT[I])(x,p,q) = 1  \LRW (g\cap \STRUCT[I])(((x,p),1),q) = 1 \LRW
     f(x,p) = 1 \]
proving $(\STRUCT[I],\SETS) \models Z := \mathrm{FC}(f,p)$.

For the proof of $\mathrm{op}$ take any operator $f := \mathrm{op}(\ARR{g})$ of $\GA$ and
any $\ARR{g} \in \FN$. For each function
$g_i$ obtain from the characterization of $\FN$ 
the functions $g_i'',g_i' \in \ARR{f}$ and parameter $p_i \in \STRUCT[I]$ s.t. \UNFIN
$g'_i(x,c,p_i) = \mu_{y {<} c}[g''_i((x,y),p_i) = 1]$.
We wish to find an $f \in \FN$ such that
$(\STRUCT[I],\SETS) \models f:=\mathrm{op}(\ARR{g})$. The idea of the proof is the same for each kind of the operator. 
We define a function $f' \in \ARR{f}$  from functions related to $g_i'$ by the operator of the same kind.
The function $f'$ ships the parameters
$\ARR{p} := p_1,\ldots,p_n$ to the functions $g_i'$. \UNFIN restriction

When the operator is $\mu$ 
we obtain $h_1, f'' \in \ARR{f}$ such that 
\begin{align*}
& \STRUCT[N] \models 
  \forall y{,}x{,}q\ h_1(y,x,q) \stackrel{\text{\ref{cl:char:term:1}}}{=} g_1''((y,x),q) \land {} \\
 & \begin{aligned} \STRUCT[N]  \models  \forall w{,}q 
     \bigl(f''(w,q) = 1 \stackrel{\text{\ref{cl:char:form:1}}}{\LRW} {} &  w = (0,0) \lor {} \\
        & \exists b{,}x{,}y{<}w\bigl(w = ((b,x),y)\land y = f_{\mu(h_1)}(b,x,q)\bigr) \ . 
   \end{aligned}
\end{align*}
\UNFIN use characterization of $\FN$ backwards 
such that for all 
$c \in \STRUCT\setminus\STRUCT[I]$   we have $f := \mathrm{Par}(f',c,p) \in \SETS$.

When the operator is defined by an explicit or recursive clausal definition whose
clauses form a set $C$ we can assume w.l.o.g. that
ll first-order variables occurring in $C$ are $\ARR{z}$
Let $q_0$ and $\ARR{q} := q_1,\ldots,q_n,q_{n+1}$ be new variables not occurring in $C$.

We systematically replace in the antecedents of clauses in $C$
all applications of $g_i(t) = z$ by $g_i'(t,q_0,q_i) = z$.
and all applications of $f(t)=z$ by $f'(t,\ARR{q}) = z$. We thereby obtain a set of clauses $C'(q_0,\ARR{q})$.

Suppose that we contrive to obtain for $\ARR{p} := p_1,\ldots,p_n,0$ a function $f' \in \ARR{f}$
such that for all $c \in \STRUCT\setminus \STRUCT[I]$ we have
\begin{gather*}
\text{$\STRUCT[N] \models  \bigwedge C'(c,\ARR{p})$ for all $\ARR{z} \in \STRUCT[I]$}   \tag{$\dagger_1$} \\
\text{for any $x \in \STRUCT[I]$ there is a $y \in \STRUCT[I]$ s.t.
    $\STRUCT[N] \models f'(x,c,\ARR{p}) = y$} \ . \tag{$\dagger_2$}
\end{gather*}
\UNFIN We can then ``contract'' the function $f'$ by taking the set $f := \mathrm{Par}(f',(\ARR{p})) \in \SETS$ which is
in $\FN$ on account of $(\dagger_2)$.
We obviously have
$(\STRUCT[I],\SETS) \models f(x) = y$ iff $\STRUCT[N] \models f'(x,b,\ARR{p}) = y$ for all $x,y \in \STRUCT[I]$ 
and $c \in \STRUCT\setminus \STRUCT[I]$.

Let $\phi$ be any atomic formula in $C$ and
$\phi'(\ARR{q})$ its corresponding formula in $C'(q_0,\ARR{q})$. We have
$(\STRUCT[I],\SETS) \models \phi$ iff $\STRUCT[N] \models \phi'(c,\ARR{p})$ for all $\ARR{z} \in \STRUCT[I]$ and 
$c \in \STRUCT \setminus \STRUCT[I]$. This
equivalence extends to all boolean combinations $\psi$ of atomic formulas in $C$ with
the corresponding combinations of $\psi'(q_0,\ARR{q})$ in $C'$.
We thus have  $(\STRUCT[I],\SETS) \models \bigwedge C$ iff $\STRUCT[N] \models  \bigwedge C'(c,\ARR{p})$
for all $\ARR{z} \in \STRUCT[I]$ and $c \in \STRUCT\setminus \STRUCT[I]$.
\UNFIN We now obtain the desired $f \in \SETS$
$(\STRUCT[I],\SETS) \models \forall \ARR{z}\bigwedge C$ from $(\dagger_1)$.

It remains to satisfy the conditions $(\dagger_i)$. For that we consider
two cases.

The case~(i). The clausal definition of $f$ is explicit.
Add to $C'(q_0,\ARR{q})$ new ``default'' clauses
$f'(x,0) = 0$, $f'(x,q_0,0) = 0$, $f'(x,q_0,q_1,0) = 0$, \ldots, $f'(x,q_1,\ldots,q_{n-1},0) = 0$.
Transform the clauses into a strict form and obtain an explicit clausal definition of
$f'$. We have
$f' \in \ARR{f}$ by \autoref{cl:ara1:cl:explicit}. The strict clauses are satisfied in $\STRUCT[N]$ from which we get
$\STRUCT[N] \models \forall \ARR{z} \bigwedge C'(c,\ARR{p})$ for all $c \in \STRUCT\setminus \STRUCT[I]$.
\UNFIN
The condition $(\dagger_1)$ is a special case of this.
The condition $(\dagger_2)$ follows from that all $g_i'(t,c,p_i)$ yield values
in $\STRUCT[I]$ when applied to such arguments.

The case~(ii). The clausal definition of $f$ is recursive.
The clauses in $C$ are parameterized
as required by the conditions in \autoref{par:rec:claus:def}.
Thus all functions $g_i'$ are applied in the antecedents of clauses in $C'(q_0,\ARR{q})$ as $g_i'((t,p),q_0,q_i)$ and
the recursive applications are $f'((t,p),q_0,\ARR{q}) = z$. The consequents of all clauses are
$f'(x,q_0,\ARR{q}) = y$.
By \autoref{cl:char:term:1} define
$h_i \in \ARR{f}$ s.t. $\STRUCT[N] \models \forall w,p,\ARR{q}\ h_i(w,p,q_0,\ARR{q}) = g_i((w,p),q_0,q_i)$.
Obtain $h' \in \ARR{f}$ such that $\STRUCT[N] \models h' := \mathrm{op}(\ARR{h})$

\UNFIN
Define $f'((v,p),q_0,\ARR{q}) = h'(v,p,q_0,\ARR{q})$
Substitute into the definition clauses for $h'$ the pair $(p,\ARR{q})$ for $p$ and
replace every $h_i(t,p,q_0,\ARR{q}) = z$ by $g_i'((t,p),q_0,q_i) = z$ and $h'(t,p,q_0,\ARR{q}) = z$ by
$f'((t,p),q_0,\ARR{q}) = z$.
It should be clear we have $\STRUCT[N] \models \forall \ARR{z},q_0,\ARR{q} \bigwedge C'(\ARR{q})$
and the condition $(\dagger_1)$ is satisfied as a special case.

\UNFIN $(\dagger_2)$ will need for any $c \in \STRUCT \setminus \STRUCT[I]$ the induction on $v$ with the formula
$f'((v,p),c,\ARR{p}) < c$  performed in $\STRUCT[N]$.

\UNFIN
mention measure. Special case when
$n=0$ and  when some  $g_i$ is applied in the form: $g_i(p)$.

\end{proof}
}

\begin{LEMMA}\label{cl:first:second:model:ara}
If $(\STRUCT, X, \ARR{f}) \models \ARITHM1{\GA}$ and
$\STRUCT[I]$ is a proper initial segment of $\STRUCT$ such that
$(\STRUCT[I], X \cap \STRUCT[I],\{f_d \cap \STRUCT[I]\}_{d \in \N}) \models \ARITHM1{\GA}$
then  there is a class $\SETS := \mathrm{FC}_{\STRUCT[I]}(\STRUCT,X,\ARR{f})$ of subsets of $\STRUCT$
such that 
$(\STRUCT[I],\SETS) \models \ARITHM2{\GA}$
and
$X \cap \STRUCT[I] \in \SETS$ as well as $f \cap \STRUCT[I] \in \FN^{(\STRUCT[I],\SETS)}$ for all $f \in \ARR{f}$.
\end{LEMMA}
\begin{proof}
Assumptions of this lemma are too weak and do not force the closure of $\SETS$ under recursive $\mathrm{op}$s of the algebra
$\ARITHM1{\GA}$, although they are sufficient for the closure under $\mu$ and explicit $\mathrm{op}$s. A strengthenning of
the assumptions to semiregular cuts is too strong, 
it forces $\SETS$ to be closed under primitive recursion. We are currently looking
into some intermediate assumptions and think that we know how to formulate them. The basic problem is that we have to know more
about the recursive operators of the algebra.
\end{proof}
\PAR{Function Comprehension and Standard Systems}
The class denoted in the preceding lemma as
$\mathrm{FC}_{\STRUCT[I]}(\STRUCT,X,\ARR{f})$
is closely related to the standard system
\[\mathrm{SSy}_{\STRUCT[I]}(\STRUCT) =
    \{\,\{\, x \mid x \in \STRUCT[I]\pAND,
                    \STRUCT \models x \in_{\mathrm{Ack}} y \,\}
        \mid y \in \STRUCT
    \,\}.
\]
There is the characteristic function of the $\Delta_0$ formula
$x \in_{\mathrm{Ack}} y$ among~$\ARR{f}$.
It produces every set from $\mathrm{SSy}_{\STRUCT[I]}(\STRUCT)$
via function comprehension,
hence $\mathrm{SSy}_{\STRUCT[I]}(\STRUCT) \subseteq
         \mathrm{FC}_{\STRUCT[I]}(\STRUCT,X,\ARR{f})$.

If additionally $\STRUCT[I] <^\STRUCT a \in \STRUCT$
and $(\STRUCT,X,\ARR{f}) \models \mathrm{exp}$,
then every $\Delta_0$~set of elements less than~$a$ is coded in~$\STRUCT$,
i.e., for every $\Delta^0_0$ formula~$\varphi(x,\ARR{z})$
we have
$(\STRUCT,X,\ARR{f}) \models \forall\ARR{z}\,\exists y\,\allowbreak
    \forall x{<}a\,( \varphi(x,\ARR{z}) \LRW x  \in_{\mathrm{Ack}} y)$
by a folklore lemma \cite[Lemma IV.2.12]{hajek-pudlak:meta}%
\cite[Prop.\ppsp2.1]{enayat-wong}.
For inductive $X \subseteq \STRUCT$,
the lemma applies to $\Delta^0_0(X,\ARR{f})$ formulas as well,
hence, in particular, to $f_d(x,p) = 1$.
Therefore,
$\mathrm{FC}_{\STRUCT[I]}(\STRUCT,X,\ARR{f}) \subseteq
    \mathrm{SSy}_{\STRUCT[I]}(\STRUCT)$.

Thus, if $\ARITHM1{\GA} \vdash \mathrm{exp}$, we have
$\mathrm{FC}_{\STRUCT[I]}(\STRUCT,X,\ARR{f}) = \mathrm{SSy}_{\STRUCT[I]}(\STRUCT)$
for every proper initial submodel
$(\STRUCT[I],X\cap\STRUCT[I],\ARR{f}\cap\STRUCT[I])$
of any $(\STRUCT,X,\ARR{f}) \models \ARITHM1{\GA}$.

%
%

\begin{THEOREM}\label{cl:first:second:conservative}
If $\GA$ is a class of function algebras
with polynomially bounded functions,
then the second-order arithmetic $\ARITHM2{\GA}$
is a lifted $\Pi_2^0$ conservative extension
of the first-order arithmetic $\ARITHM1{\GA}$.
\end{THEOREM}

\begin{proof}
    Take any~$\GA$ with polynomially bounded functions.
    $\ARITHM2{\GA}$
    is a lifted extension
    of $\ARITHM1{\GA}$ by the previous theorem.

    For lifted $\Pi^0_2$ conservativity,
    suppose a $\Pi^0_2$ sentence~$\forall x\exists y\varphi(x,y)$
    is not provable in $\ARITHM1{\GA}$.
    We need to show that its lifted form is not provable in $\ARITHM2{\GA}$,
    i.e., to find a model of $\ARITHM2{\GA}$
    with sets~$X$ and~$\smash{\ARR{f}_k}$
    satisfying~$\Phi_\varphi = \Phi_k$
    while not satisfying $\forall x\,\exists y\,\varphi(x,y)$.

    Since $\ARITHM1{\GA} \not\vdash \forall x\exists y\varphi(x,y)$,
    the theory $\ARITHM1{\GA} + \forall y\,\varphi(e,y)$
    with $e$~a new constant is consistent.
    Moreover, for another new constant~$c$ every finite subset of the theory
    \[
        T \coloneq \ARITHM1{\GA} \cup \{ \forall y\,\varphi(e,y) \} \cup
            \{\, e^k <\nobreak c \mid k \in \N \,\}
    \]
    is consistent as well.
    By compactness, $T$~is consistent, and by completeness,
    it has a model $\STRUCT[N] \coloneq (\STRUCT,X,\ARR{f},e,c)$.

    Let $\STRUCT[I]$ be the $\LANG_1$ structure with the domain
    $I = \{\, x \mid x \in \STRUCT\pAND, \STRUCT \models x < e^k\pAND,
            k \in \N \,\}$.
    $\STRUCT[I]$~is a proper initial segment of $\STRUCT$
    due to the definition of~$I$ and since $I <^\STRUCT c$.
    Since the functions of $\GA$ are polynomially bounded,
    $I$~is closed under the functions~$\ARR{f}$.
    Moreover, $e \in I$ and $\forall y\,\varphi(e,y)$ is~$\Pi^0_1$,
    and thus absolute.
    Hence $(\STRUCT[I], X \cap \STRUCT[I], \{f_d \cap \STRUCT[I]\}_{d \in \N}) \models
            \ARITHM1{\GA} + \forall y\,\varphi(e,y)$.

    \autoref{cl:first:second:model:ara}
    now gives us a system of sets~$\SETS$
    such that $(\STRUCT[I],\SETS) \models \ARITHM2{\GA}$
    with $\{X \cap \STRUCT[I]\} \cup \{f_d \cap \STRUCT[I]\}_{d \in \N} \subseteq \SETS$.
    We thus have $(\STRUCT[I],\SETS) \models
    \Phi_k(X \cap \STRUCT[I], \{f_d \cap \STRUCT[I]\}_{d<k})$,
    and $(\STRUCT[I],\SETS) \models \forall y\,\varphi(e,y)\,
            (X \cap \STRUCT[I],\allowbreak \{f_d \cap \STRUCT[I]\}_{d<k})$.
    Hence $(\STRUCT[I],\SETS) \not\models \varphi{\uparrow}$,
    as desired.
    \qed
\end{proof}

%
%


\section{Some Function Algebras for Complexity Classes}\label{sc:complexity}
In this paper we have introduced a general framework for connecting the provable functions of first and second-order
recursive arithmetics.
This section serves as an illustrative application where we formulate several subelementary recursive arithmetics
capturing some of the main complexity classes. For this reason the section does not contain any theorems and its assertions are mostly
only sketched out.

\PAR{Space Algebras $\SA$}\label{par:bounded:recursion}
The function operator $f:=\mathrm{PR}(g,h)$ (see \autoref{par:opers}) is \emph{bounded} when for all $x \in \N$ we have
$f(x) \leq b(x)$.
This is a semantic condition and so we cannot use it as an operator (it does not always yield a function).
We define instead
the operator $f := \mathrm{BPR}(g,h)$ of \emph{bounded primitive recursion} by the following clausal
definition with the identity function as measure:
\begin{gather*}
   g(p) = z \land z \leq p \RW f(0,p) = z \\
   h((v,f(v,p)),p) = z \land z \leq p \RW f(v+1,p) = z \ .
\end{gather*}
When we present a non-strict clausal definition like this we trust the reader that they can transform it
into a strict one. In this case
this means the applications of functions in quasi-terms must be unnested (i.e. $f(v,p)$), The consequents have
to be brought to the strict
form $f(x) = y$ which involves possibly renaming variables and
moving the terms in the arguments of $f$ in the consequents (such terms are called in computer programming
\emph{patterns}) by moving them into antecedents.
For instance, in the first clause we put $x = (u,p) \land u = 0$ into the antecedent and add the missing clauses, so called
\emph{default} clauses when $f$ will yield $0$, i.e. $x = (u,p) \land u \not = 0 \land \cdots$, or the clause $f(0) = 0$.
After adding the default clauses, the clauses should be conjuncted into one formula and its first-order variables universally closed.

We designate by $\SA$ the class of \emph{space algebras} obtained by adjoining the operator $\mathrm{BPR}$ to the operators of
the class $\DA$. The function $f$ yielded by the operator $\mathrm{BPR}$ is non-growing and therefore
the functions of the algebra are polynomially bounded (note that the parameter
$p$ of bounded primitive recursion can be set to at most a polynomial in $x$).
We will see below that the algebra $\SA$ is suitable for the characterization of space complexity classes.

It should be clear that $\SA$ is closed under primitive recursion $f := \mathrm{PR}(g,h)$ which is bounded by $b$ because
we can define $f_1 := \mathrm{BPR}(g,h)$ and then $f(x) = f_1(x,\max(b(x),1))$.

The computation of $f(x,p)$ defined by $\mathrm{BPR}$ when done by iteration requires the space sufficient
to hold two numbers no larger than $p$ provided $x \leq p$. This is the way computations in space complexity classes are done.

\PAR{Time Algebras $\TA$}\label{par:bounded:nested:recursion}
We define the operator of \emph{special nested recursion}
$f := \mathrm{SNR}(g,h)$ by the following schema of clausal definitions:
\begin{gather*}
g(x,p) = (0,z) \land z < x \land f(z,p) = v \land h((x,v),p) = w \land w < x \RW f(x,p) = f(w,p) \\
g(x,p) = (1,z) \land z \leq p \RW f(x,p) = z \ .
\end{gather*}
The function $f$ is obviously non-growing and the identity function is its measure.

We designate by $\TA$ the class of \emph{time algebras} obtained by adjoining the operator $\mathrm{SNR}$ to the operators of $\DA$.
The functions of $\TA$ are polynomially bounded because the parameter
$p$ of special nested recursion can be set to at most a polynomial in $x$.

We call a recursive clausal definition of $f$ \emph{bounded nested} if identity is its measure function
and for some bounding function $p$ we have $f(v) \leq p(v)$ for all
$v \in \N$. We will now show the algebras $\TA$ closed under such definitions by defining $f$ by special nested recursion.
We first transform the definition of $f$ to the explicit function $h$
as in \autoref{par:rec:claus:red}
from where we also obtain the constant $J$ giving the maximal nesting of recursions in the clauses of $f$.
It basically remains to reduce $J$ to $2$.
To that end we define $f_1 :=\mathrm{SNR}(g_1,h_1)$
where the auxiliary functions have the following explicit clausal definitions:
\begin{gather*}
 p = (m,b,p') \land v = [x,c']_b \land  m \dotminus c' = c \land h(x,c) = (0,z) \RW g_1(v,p) = (0,[z,m]_b) \\
 p = (m,b,p') \land v = [x,c']_b \land  m \dotminus c' = c \land h(x,c) = (1,z) \RW g_1(v,p) = (1,z) \\
 \\
   p = (m,b,p') \land v = [x,c']_b \land  m \dotminus c' = c  \RW   h_1((v,w),p) = [x,m\dotminus(c\oplus_m (w,0))]_b \ .
\end{gather*}
The argument $v$ in both functions is a pair of numbers $x$ and $c'$ which is not coded by the Cantor's function but rather
as two digits of a number in the base $b$: $[x,y]_b := x \cdot b + y$ which has the pairing property when $x,y < b$.
This function, both
of its projections, as well as
the \emph{modified subtraction} $\dotminus$ (yielding $0$ if the result should be negative) are easily derivable in $\DA$.

The lists $c$ passed to the function $h$ contain at most the values $p:= p(x)$
and they grow during
the computation of $f_1$ from the length of $0$ to the maximal length $J$.
They are thus at most $m := (\overbrace{p,\ldots,p}^{J},0)$ large.
We now have an explicit definition of $f$ as
$f(x) = f_1([x,m]_b,m,b,b^2)$
where $b := \max(x,m)+1$. Note that
the lower $b$-digit $c'$ of $v$ codes $c$ ``backwards'' where the list is $c:= m \dotminus c'$.
This makes the measure of $f_1$ the identity function.

It remains to derive the \emph{bounded} list concatenation $x\oplus_m y$ (abbreviating ${\oplus}((x,y),m)$)
as a non-growing function.
Note that we do not have the general concatenation function in $\TA$ because it is bounded by an exponential
with the exponent depending on $L(x)$. Since $L(c) \leq J$ it suffices to use the following
approximation explicitly defined in $\DA$:
\begin{gather*}
     0 \oplus_m y = y \\
   (z_1,y) = z \land z \leq m \RW (z_1,0) \oplus_m y = z \\
   \vdots \\
   (z_1,\ldots ,z_{J-1},y) = z \land z \leq m \RW (z_1,\ldots,z_{J-1},0) \oplus_m y = z \ .
\end{gather*}
Note that the operator $f:= \mathrm{BPR}(g,h)$ of bounded primitive recursion yields a non-growing function and
so its definition is a bounded nested one and we have $\SA \subseteq \TA$.

As we have seen in \autoref{par:rec:claus:red},
a straightforward evaluation of nested recursive definition of $f(x)$ with the identity as measure
needs a stack whose length is $x$ and
time (length of iteration) exponential in $x$.
We can reduce the length of iteration to $x$ if the function is bounded by $p(x)$ because we can encode the values
$f(x-1), f(x-2), \ldots, f(1), f(0)$ as $x$ digits of a number $s$ in the base $b := p(x)+1$ and compute the number $f(x)\cdot b^x + s$
by looking up the recursive applications in the definition of $f$ as digits of $s$. Thus the computation of $f(x)$ can be done by
\emph{course-of-values} recursion (see e.g. \cite{rose}) requiring the time $x$ and space sufficient for the course-of-values
sequences $s < p(x)^{x+1}$. This is how computations in time complexity classes are done.
To our best knowledge it was Jones in \cite{jones} who has noticed that the exponentially many steps of the stack computation
can be reduced to $x$ steps  by using a look-up table of already computed function values. Such techniques are
called in computer programming \emph{memoization}.
However, the combination of a stack with a look-up table is not necessary, 
because the course-of-values recursion does the trick directly.

\PAR{Classes of Computational Complexity and Function Algebras}\label{par:comp:classes}

We wish to connect the classes of computational complexity with the classes of function algebras $\GA$.
Toward that goal we present the complexity classes $\cal C$ in two forms.
A \emph{type-0} class is a set of subsets of $\N$.
For a set $P$ in such a class we decide whether or not
$x \in P$ by presenting $x$ to a computing device (usually a Turing machine) in the binary representation as
finite sequences of $0$ and $1$.
A \emph{type-1} class is a set of subsets of $2^{{<}\N}$. For a set $P$ in such a class we decide whether or not
$X \in P$ by presenting to a computing device
the binary representation
of Ackermann's encoding of $X$, i.e.a finite sequence $x_{\PS{X}-1},\ldots,x_0$ such that for all $i < \PS{X}$ we have $x_i =1$ if
$i \in X$ and $0$ otherwise.

A type-0 class $\cal C$ is \emph{$0$-characterized by the class $\GA$} if
$\cal C$ is the set of $P \subseteq \N$ such that
there is a provably recursive
$(0{-}1)$ valued function $f$ of $\ARITHM1{\GA}$ and $P = \{ x \in \N \mid f(x,\emptyset) = 1\}$.

A type-1 class $\cal C$ is \emph{$1$-characterized by the class $\GA$} if
$\cal C$ is the set of  $P \subseteq 2^{{<}N}$ such that
there is a provably recursive
$(0{-}1)$ valued function $f$ of $\ARITHM1{\GA}$ and $P = \{ X \in 2^{{<}N} \mid f(\PS{X},X) = 1 \}$.

The type 0 and 1 characterizations of complexity classes by means of different
inputs to function algebras stem from the second author's
cooperation with L.~Kristiansen (see e.g. \cite{kv}).

%
%

\PAR{Some Function Algebras Characterizing Complexity Classes}\label{par:comp:classes:examp}

The space class of algebras $\SA$ $0$-characterizes the complexity class $\mathrm{LINSPACE}$ (i.e. $\SPACE({\cal O}(n))$).
The same class $1$-characterizes the class $\mathrm{LOGSPACE}$ (i.e. $\SPACE({\cal O}(\log(n)))$).
The $0$-characterization comes from
the early result of Ritchie \cite{ritchie,clote} that $\mathrm{LINSPACE}$ is
identical to the predicates of the Grzegorczyk's class ${\cal E}^2_\ast$ \cite{grzeg,rose}.
The class ${\cal E}^2$ is defined by bounded primitive recursion and so are the algebras $\SA$. For a more detailed discussion
see \cite{kv}.

The time class of algebras $\TA$ $0$-characterizes the complexity class $\mathrm{ETIME}$ (i.e. $\TIME(2^{{\cal O}(n)})$).
The same class $1$-characterizes the class $\mathrm{PTIME}$ (i.e. $\TIME(n^{{\cal O}(1)})$).
The characterization of $\mathrm{PTIME}$ by bounded nested recursion
is from \cite{jones}, the modification to the $0$-characterization is obvious
because
of exponentially more time available (as a function of input): $x$ vs. $\PS{X}$.
There is an old characterization of $\mathrm{ETIME}$ by
bounded twofold recursion by Monien \cite{monien,clote} which is, however, not nested.

The class of algebras $\DA$ $0$-characterizes the complexity class $\mathrm{LINTH}$ (linear time hierarchy) which is the
class of $\Delta_0$-definable predicates (see \cite{clote}).
The same class $1$-characterizes the class $\mathrm{LOGTH}$ (log time hierarchy), also known as $\mathrm{FOL}$ (first-order logic)
in finite model theory.

If we add the nullary operator $f := \mathrm{E}$ (see \autoref{par:opers}) to the operators of $\DA$
we obtain the class of algebras $\DEA$ which
are obviously the algebras of elementary functions with oracles.
We can define a subexponential  operator $f := \#$ yielding the function $f(x) = 2^{\PS{x}^2}$ which has
the same growth rate as the \emph{smash} function $x \mathrel{\#} y = 2^{\PS{x}\cdot\PS{y}}$ or the function
$\omega_1(x,y) = x^{\PS{y}}$ of Wilkie and Paris \cite{wilkie-paris} where
$\PS{x}$ is the size of $x$ in binary representation. Adding the operator to $\DA$ gets the class $\DSA$ which
$0$-characterizes the complexity class $\mathrm{PH}$ of \emph{polynomial time hierarchy} (see c.f. \cite{hajek-pudlak:meta}).
Adding the $\#$ operator to the algebra $\SA$ gets the class $\SSA$ which $0$-characterizes the complexity class
$\mathrm{PSPACE}$ (i.e. $\SPACE(n^{{\cal O}(1)})$) (see \cite{clote}).

Weak K\"onig lemma ($\WKL^-$) does not seem to be directly usable with the characterization
of non-deterministic classes like $\mathrm{NP}$ (non-deterministic polynomial time), or
$\mathrm{NL}$  (nondeterministic log space) because the lemma deals with infinite trees.

There are the following well-known inclusions of the complexity classes
\[\mathrm{LOGSPACE} \subseteq \mathrm{LINTH} \subseteq \mathrm{LINSPACE} \subseteq \mathrm{ETIME} \]
 and
\[ \mathrm{LOGTH} \subseteq \mathrm{LOGSPACE} \subseteq \mathrm{PTIME} \subseteq \mathrm{PH} \subseteq \mathrm{PSPACE} \ . \]
Frustratingly, the questions whether any of the inclusions are
strict are the major open problems of computational complexity, although we have
$\mathrm{LOGSPACE} \subsetneq \mathrm{LINSPACE} \subsetneq \mathrm{PSPACE}$.

The arithmetic $\ARITHM1{\DEA}$ is obviously equivalent to the Elementary function arithmetic $\mathrm{EFA}$ and
$\ARITHM1{\DSA}$ is a conservative extension of the theory $I\Delta_0(\Omega_1)$ of \cite{wilkie-paris} where
$\Omega_1$ states that the function $\omega_1$ is total. Although it is known
that the hierarchy $I\Delta_0(\Omega_k)$ is strict and spans the theory $\Delta_0(\mathrm{exp})$ (see, e.g., \cite{hajek-pudlak:meta}),
its levels $k>1$ are not directly connected to any major complexity classes.

\section{Conclusions and Future Work}
For the final version of this paper we plan
to tidy up the axioms in $\BASIC$.
We think that the languages of arithmetic integrating the fours forms of its presentation (by induction on first-order formulas,
by recursive arithmetics (both first- and second order), and by second order arithmetics in the style of Friedman and Simpson) should
be based on the pairing function as the basic binary function.
The language of arithmetics should contain the constant $0$, the symbol $({\cdot},{\cdot})$ of pairing, and
possibly the relation
symbol $<$ as basic. All remaining symbols should be set constants (in the first-order theories) and set variables (in the
second-order theories).
The axiomatization could be by the pairing axioms $\mathrm{P}1{-}2\mathrm{2}$ as well the quasi-formulas
characterizing the successor function with the help of the modified Cantor's pairing:
\[ S(0) = (0,0) \quad S(x,S(y)) = (S(x),y) \quad  S(x,0) = (0,S(x)) \ . \]
From this we get the usual properties of the successor function.

The properties of the set constants ${+}$ and ${\cdot}$, which are in this draft designated by $\oplus$ and $\otimes$
respectively, can
be axiomatized by recursive quasi-formulas. We can possibly replace the relation symbol $<$ by the set constant
$<_\ast$ denoting its characteristic function. For syntactic comfort we should
use quasi-terms and quasi-formulas as abbreviations
for their unnested forms. We did not systematically use them in this draft because we have in our
arithmetic languages the symbols $S$, ${+}$, ${\cdot}$, and $<$ available as the standard ones.

The recursive arithmetics introduced in this draft are formulated  in such a way that we can add to them as initial functions
the hierarchy functions of Grzegorczyk's hierarchy to characterize the theory $\IND{\Sigma}_1$ and the functions of
the Weiner-Schwichtenberg
hierarchy for the characterization of fragments $\IND{\Sigma}_{i+2}$ of $\mathrm{PA}$ (cf.~\cite{avigad-sommer}).

For the second-draft of this paper we should present the recursive
arithmetics as triples $\ARITHM1{\GA}$,  $\ARITHM2{\GA}$, and $\ARITHM2{\GA}{+}\WKL$. We think that the last theory can
be characterized similarly as the theory $\WKL^-_0$ vs. $\RCA^-$. The missing element, completing this to a quadruple of the kind
discussed in the introduction, is a theory with induction, say $\IND{\Delta_0}(e)$ with $e$ an axiom asserting the totality
of some subexponential function.
It seems that the space and time arithmetics
$\ARITHM1{\SA}$,  $\ARITHM2{\TA}$
cannot be fully characterized in this way. For instance, 
the smash function $2^{\PS{x}^2}$ does not seem to work except in the
cases mentioned in \autoref{par:comp:classes:examp}.

The obvious area for research is the characterization of major non-deterministic classes ($\mathrm{NPTIME}$, $\mathrm{NLOGSPACE}$)
by means of recursive arithmetics. Although non-determinism can be viewed as
a search for a path  in a tree expressing a particular property, the approach through
$\WKL$ does not seem to work because of the lack of exponentiation
(not too many definable trees) and it probably  will not be possible to downscale the infinite trees to the finite ones of complexity
theory.

Another area for research is the characterization of subexponential second-order models in the form
$(\STRUCT, \mathrm{SSy}(\STRUCT))$. We were not able to do this because there do not seem to be sufficiently many coded sets.


\begin{thebibliography}{1}
\bibitem{avigad-sommer}
    Avigad,~J. and Sommer,~R.:
    \emph{The Model-Theoretic Ordinal Analysis of Predicative Theories.}
    JSL vol. 64:327--349, 1999
\bibitem{beklemishev}
 Beklemishev, L.:
     \emph{On the Induction Schema for Decidable Predicates.} JSL vol. 68, 2003.
\bibitem{clote}
  Clote, P.: \emph{Computation Models and Function Algebras.} in Handbook of Computability Theory. Elsevier 1999.
\bibitem{cook-kolokolova:poly}
  Cook, S. and Kolokolova, A.:
  \emph{A Second-Order System for Polytime Reasoning Based on Gr\"adel's Theorem.}
    APAL vol. 124, 2003.
\bibitem{ebbinghaus}
  Ebbinghaus,  H-D. and Flum, J. \emph{Finite Model Theory.} Springer Science, 2005.
\bibitem{enayat-wong}
  Enayat, A. \ and Wong, T. L.:
  \emph{Unifying the Model Theory of First-Order and Second-Order Arithmetic.} APAL vol. 168, 2017.
\bibitem{grzeg}
   Grzegorczyk, A.: \emph{Some Classes of Recursive Functions.} Rozprawy Matematyczne Vol. 4, 1953
\bibitem{gurevich}
  Gurevich, Y.: \emph{Algebras of Feasible Functions.} IEEE Found. of Comp. Sci. Symp. vol. 24, 1983.
\bibitem{ferreira}
  Ferreira, F.:
  \emph{A Simple Proof of Parsons' Theorem.}
  Notre Dame Journal of Formal Logic, 2003.
\bibitem{hajek-pudlak:meta}
  Hájek, P. and Pudlák, P.:
  \emph{Metamathematics of First-Order Arithmetic.} Springer, 1993.
\bibitem{jones}
 Jones, N.:
  \emph{Computability and Complexity From a Programming Perspective.}
  The MIT Press, 1997.
\bibitem{kaye}
  Kaye, R., \emph{Models of Peano Arithmetic.} Oxford Logic Guides, Clarendon Press, 1991.
\bibitem{kossak:schmerl}
  Kossak, R. and Schmerl, J.: \emph{The Structure of Models of Peano Arithmetic.} Oxford Logic Guides, Clarendon Press, 2006.
\bibitem{kpt}
  Krajíček, J., Pudlák, P., and Takeuti, G.:
  \emph{Bounded arithmetic and the polynomial hierarchy.}
  APAL vol. 52:143--153, 1991.
\bibitem{kv}
  Kristiansen, L. and Voda, P. J.:
  \emph{Programming languages capturing complexity classes.} Nordic Journal of Computing vol. 12, 2005.
\bibitem{monien}
  Monien, B.: \emph{A Recursive and Grammatical Characterization of Exponential Time Languages.} Theoret. Comp. Sci. vol. 3, 1977
\bibitem{ritchie}
  Ritchie, R. W.: \emph{Classes of Predictably Computable Functions.} Trans. Amer. Math. Soc. vol. 106, 1963.
\bibitem{rose}
  Rose, H. E.: \emph{Subrecursion Functions and Hierarchies.} Oxford Logic Guides, Clarendon Press, 1984.
\bibitem{shoenfield}
  Shoenfield, J. R.: \emph{Mathematical Logic}. Association for Symbolic Logic, 1967.
\bibitem{simpson:subsystems}
  Simpson, S.G.:
  \emph{Subsystems of Second-Order Arithmetic.} Cambridge University Press, 2009.
\bibitem{simpson-smith:factorization}
   Simpson, S.G.\ and Smith, R.L.:
  \emph{Factorization of Polynomials and $\Sigma^0_1$ Induction.} APAL vol. 21, 1986.
\bibitem{slaman}
  Slaman, T.A.:
  \emph{$\Sigma_n$-Bounding and $\Delta_n$-Induction.}  Proc. Amer. Math. Soc. vol. 132, 2004.
\bibitem{voda:pa-and-cl}
  Voda, P.J.: \emph{Peano Arithmetic and Clausal Language.}
  Lecture notes. Bratislava: Comenius University, 2004.
  [online] \url{http://dai.fmph.uniba.sk/~voda/pa.pdf}.
\bibitem{wilkie-paris}
  Wilkie, A.J. and Paris J.B.: \emph{On the Scheme of Induction for Bounded Arithmetic Formulas}. APAL vol. 35, 1987.
\bibitem{zambella:poly}
  Zambella, D.:
  \emph{Notes on Polynomially-Bounded Arithmetic.} JSL vol. 61, 1996.
\end{thebibliography}
\end{document}